\documentclass[superscriptaddress,iop,apj]{emulateapj}
\usepackage{latexsym}
\usepackage{amssymb}
\usepackage{amsfonts}
\usepackage{amsmath}
\usepackage{bm}
\usepackage[normalem]{ulem}
\usepackage{multirow}
\usepackage{natbib} 
\usepackage{subfigure}
\usepackage{color}
\usepackage{calc}
\usepackage{amsmath,amssymb,graphicx}
\usepackage{amssymb,amsmath}
\usepackage{tensor}
\usepackage{bm}
\usepackage{microtype}
\usepackage{booktabs}
\usepackage{times}
\usepackage[varg]{txfonts}
\usepackage[colorlinks, pdfborder={0 0 0}]{hyperref}
\usepackage{subfigure}
\usepackage{lipsum}
\usepackage{breakurl}
\usepackage{graphicx}
\usepackage{pbox}

\definecolor{LinkColor}{rgb}{0.75, 0, 0}
\definecolor{CiteColor}{rgb}{0, 0.5, 0.5}
\definecolor{UrlColor}{rgb}{0, 0, 0.75}
\definecolor{darkgreen}{rgb}{0,0.42,0.24}
\hypersetup{linkcolor=LinkColor}
\hypersetup{citecolor=CiteColor}
\hypersetup{urlcolor=UrlColor}
\allowdisplaybreaks
\DeclareFontFamily{OT1}{pzc}{}
\DeclareFontShape{OT1}{pzc}{m}{it}{<-> s * [1.10] pzcmi7t}{}
\DeclareMathAlphabet{\mathpzc}{OT1}{pzc}{m}{it}

\newcommand{\be}{\begin{equation}}
\newcommand{\ee}{\end{equation}}
\newcommand{\ber}{\begin{eqnarray}}
\newcommand{\eer}{\end{eqnarray}}
\newcommand{\psr}{PSR~J1023+0038}
\newcommand{\xss}{XSS~J12270$-$4859}
\newcommand{\igr}{IGR~J18245$-$2452}
\newcommand{\rxs}{1RXS~J154439.4$-$112820}

\def\bea{\begin{eqnarray}}
\def\eea{\end{eqnarray}}

\begin{document}
\title{Timing Observations of PSR~J1023+0038 During a Low-Mass X-ray Binary State}
\shortauthors{Jaodand et al.}
\author{Amruta Jaodand\altaffilmark{1,2,*}, Anne M. Archibald\altaffilmark{1,2}, Jason W.~T. Hessels\altaffilmark{1,2}, Slavko Bogdanov\altaffilmark{3}, Caroline R. D'Angelo\altaffilmark{4}, \\ Alessandro Patruno\altaffilmark{4,1}, Cees Bassa\altaffilmark{1} \& Adam T. Deller\altaffilmark{1}}
\altaffiltext{1}{ASTRON, the Netherlands Institute for Radio Astronomy, Postbus 2, 7990 AA, Dwingeloo, The Netherlands}
\altaffiltext{2}{Anton Pannekoek Institute for Astronomy, University of Amsterdam, Science Park 904, 1098 XH, Amsterdam, The Netherlands}
\altaffiltext{3}{Columbia Astrophysics Laboratory, Columbia University, 550 West 120th Street, New York, NY 10027, USA}
\altaffiltext{4}{Leiden Observatory, Leiden University, PO Box 9513, NL-2300 RA Leiden, The Netherlands}
\altaffiltext{*}{To whom correspondence should be addressed; E-mail: \protect \email{jaodand@astron.nl}.}

\begin{abstract}

Transitional millisecond pulsars (tMSPs) switch, on roughly multi-year
timescales, between rotation-powered radio millisecond pulsar (RMSP) 
and accretion-powered low-mass X-ray binary (LMXB) states.
The tMSPs have raised several questions related to the nature of 
accretion flow in their LMXB state and the mechanism that causes the
state switch. The discovery of coherent X-ray pulsations from
\psr\ (while in the LMXB state) provides us with the first opportunity to
perform timing observations and to compare the neutron star's spin 
variation during this state to the measured spin-down in the RMSP state.  
Whereas the X-ray pulsations in the LMXB state likely indicate that some material 
is accreting onto the neutron star's magnetic polar caps, radio continuum observations 
indicate the presence of an outflow. The fraction of the inflowing material being ejected 
is not clear, but it may be much larger than that reaching the neutron star's
surface. Timing observations can measure the total torque on the neutron star. 
We have phase-connected nine \textit{XMM-Newton} observations of \psr
over the last $2.5$\,years of the LMXB state to establish a precise measurement of spin 
evolution. We find that the average spin-down rate as an LMXB is $26.8\pm0.4$\% 
faster than the rate ($-2.39\times10^{-15}$\,$\rm{Hz}~\rm{s}^{-1}$) determined during the RMSP state. This shows that negative 
angular momentum contributions (dipolar magnetic braking and outflow) exceed positive 
ones (accreted material), and suggests that the pulsar 
wind continues to operate at a largely unmodified level. 
We discuss implications of this tight observational constraint in the context of possible 
accretion models.

\end{abstract}

\maketitle


\section{Introduction}

Neutron stars exist with rotation rates as high as at least $\nu_{\rm
  spin} = 716$\,Hz \citep{HRS:2006}, corresponding to a transverse
velocity at the stellar equator of ${\sim} 0.2$\,c.  The pulsar
`recycling' mechanism is the accepted scenario for understanding how
neutron stars can acquire such remarkably rapid rotation rates
\citep{ACR:1982, RS:1982}.  Millisecond rotation periods can be measured
directly in a number of neutron star system types: (1). radio
millisecond pulsars (RMSPs), where persistent and coherent radio
pulsations provide unparalleled precision on the rotational and
orbital ephemeris \citep[e.g.,][]{RSA:2014}; (2). accreting
millisecond X-ray pulsars (AMXPs), a type of low-mass X-ray binary (LMXB)
where, occasionally, channelled accretion onto the magnetic polar caps
produces coherent X-ray pulsations at the rotation rate
\citep[][and references therein]{PW:2012}; and (3). LMXB burst oscillation sources, where
oscillations corresponding approximately to the neutron star's
rotation rate are detected during Type-I X-ray bursts
\citep{CMG:2003, Anna:2012}.

Direct observational evidence for the recycling scenario has recently
come from the discovery of a population of transitional millisecond
pulsars (tMSPs)\footnote{An earlier link was also established by the
  discovery of the first AMXP, SAX~J1808.4$-$3658
  \citep{WK:1998}.}. The tMSPs are sources that switch between states
as a rotation-powered RMSP and an accretion-powered LMXB.  Three
confirmed tMSPs are known: \psr~\citep{ASR:2009},
\xss~\citep{BPH:2014}, and \igr~\citep{PFC:2013}.  One additional
candidate tMSP, \rxs, has also been proposed
\citep{BH:2015}. Although this source shows remarkably similar
  observational phenomena to the known tMSPs, it has so far only been
  observed in the LMXB state and its rotational period remains unknown.
Thus far, all three known tMSPs are eclipsing `redback' millisecond
pulsars with non-degenerate ${\sim} 0.2$\,M$_\mathrm{\odot}$ companions
\citep{Mal:2013}.  This suggests that other known redback pulsars may
also transition to accreting states\footnote{Less clear is
  whether the eclipsing RMSPs known as `black widows', which have
  $\ll\,0.1$\,M$_\mathrm{\odot}$ degenerate companions, will turn out to be
  tMSPs as well.  The canonical AMXP, SAX~J1808.4$-$3658, is
  black-widow-like, and previous authors presented evidence that it
  turns on as a rotation-powered RMSP during X-ray quiescence
  \citep[][though no radio pulsations have yet been
    observed]{BDD:2003}.  On the other hand, many known black widows in the
  RMSP state have been found to be under-filling their Roche lobe in
  at least some cases \citep[e.g.,][]{BKR:2013}, suggesting that a
  transition to active accretion is unlikely in those cases.}.
Conversely, while in their LMXB state, the radio pulsar is no longer
detected \citep[even up to relatively high radio frequencies of ${\sim}
5$\,GHz and using the 305\,m Arecibo telescope,][]{SAH:2014}, but a host of new, 
multi-wavelength observational phenomena
are seen, as we describe below.

\psr\ (hereafter J1023) is the best-studied tMSP, and
detailed observations are aided by its proximity to the Earth \citep[$d =
  1368^{+42}_{-39}$\,pc, as determined by a radio interferometric measurement of 
  geometric parallax,][]{DAB:2012}.  A long-term, radio-derived timing solution
has provided precise rotational and orbital parameters for the system
\citep{AKH:2013}.  While visible as an RMSP, J1023 shows orbital
modulation of its optical and X-ray brightness (orbital period $P_{\rm
  orb} = 4.8$\,hr). The X-ray modulation is likely the result of the
X-rays being produced in a shock near the companion's pulsar-facing
side, which is partially eclipsed during the orbit
\citep{BAH:2011}. The shock is created by the interaction of the
pulsar and companion winds, and heats the companion's face such that
the optical light curve is also modulated at the orbital period
\citep{BKR:2013, TA:2005}.

J1023 transitioned to an LMXB state in 2013 June \citep{SAB:2013, TLL:2013, PAH:2014,SAH:2014}, and
remains in this state until now (2016 May).  The state transition was signaled by the disappearance
of the radio pulsar, as well as a sudden enigmatic brightening in
$\gamma$-rays by a factor of ${\sim} 5$ \citep{SAH:2014}.  Though the
radio pulsar is no longer detectable, a variable, flat-spectrum radio 
continuum source, which is suggestive of a collimated outflow, has appeared
\citep{DMM:2015}.  Coherent X-ray pulsations have also
been detected and interpreted as originating from heating of the
magnetic polar caps by inflowing accretion material \citep{ABP:2015}.
This means that J1023 is also an AMXP, albeit at X-ray luminosities
much lower than previously observed in other AMXPs (here
$L_X {\sim} 10^{33}$\,erg\,s$^{-1}$ compared to $L_X {\sim}
10^{35-36}$\,erg\,s$^{-1}$ seen in other, more distant sources while they are in outburst).

During J1023's LMXB {\it state}, the X-ray brightness switches between
three reproducible luminosity {\it modes}\footnote{As in previous
  works \citep[e.g.,][]{BAB:2015}, we explicitly use the term `state'
  to refer to the RMSP and LMXB states of the system, whereas we used
  the term `mode' to distinguish between the three modes of X-ray
  brightness seen during the LMXB state.}: (1). high-mode ($L_X
{\sim} 10^{33}$\,erg\,s$^{-1}$), present ${\sim} 70-80$\% of the time;
(2). low-mode ($L_X {\sim} 5\times10^{32}$\,erg\,s$^{-1}$), present
${\sim} 20$\% of the time; and (3) occasional flares ($L_X {\sim}
5\times10^{34}$\,erg\,s$^{-1}$) present for about 2\% of the time.
The coherent X-ray pulsations appear only in the
high mode.  We note that \xss\ also shows a highly similar behavior with
three modes of X-ray brightness and coherent X-ray pulsations in high mode \citep{PMB:2015}.

The wealth of observational phenomena seen in J1023 provides many clues
as to the nature of the accretion in the LMXB state, though no single,
self-consistent picture has yet emerged \citep[nonetheless, see][for
  interpretations of the observed phenomena]{SAH:2014, CBC:2013,
  ABP:2015,PAH:2014, TLK:2014, TamLK:2014, PT:2015, BB:2015,
  Bed:2015}. Importantly, the aforementioned observational phenomena
seen in J1023 have also been observed in the other two known tMSPs,
\xss~and \igr, as well as the tMSP candidate \rxs.  As such, though
the tMSPs have suddenly presented many new puzzles, they have at least
presented a consistent observational picture that can be used as the
foundation for building our theoretical understanding.

A primary question is: what causes the transitions between RMSP and
LMXB states?  These occur rapidly \citep[within at most
  weeks;][]{PFC:2013, Papitto:2013, SAH:2014}, and the states themselves last for months to
years (as of this writing, J1023 has been in its current LMXB state for
close to 3 years).  The nature of the accretion during the LMXB state
is also an intriguing open question: we know that it is relatively
stable on month to year timescales but that there is likely both
inflow and outflow of material and the X-ray light curves switch
between three luminosity modes on timescales of minutes to hours \citep{PAH:2014,BAB:2015}.
These, and other related questions, hold the promise of coming to a
much deeper understanding of pulsar recycling.

Furthermore, the tMSPs may prove to be valuable laboratories for
studying accretion onto magnetized compact objects in a more general
sense.  We aim to distinguish between plausible accretion models such
as propeller mode accretion \citep{IS:1975,ST:1993}, the trapped disk
scenario \citep{SS:1975,DS:2011}, a radiatively inefficient accretion
flow model \citep{RBB:1982,CFM:2015}, etc.

A crucial contribution toward this will come from determining how the
neutron star spin changes during its LMXB state compared to its RMSP
state--where the spin-down is determined to high precision using
coherent timing of the radio pulsations.  If the observed coherent
X-ray pulsations \citep{ABP:2015} indeed come from channelled accretion onto the
magnetic poles of the neutron star, then this would in principle
induce a spin-up torque (though such a torque could be modest if
the accretion rate onto the neutron star is very small). Conversely,
the observed radio continuum emission from a collimated outflow \citep{DMM:2015}
suggests a spin-down torque due to infalling material 
being ejected by interactions with the rapidly rotating neutron star 
magnetosphere. In the RMSP state, spin-down is dominated by the pulsar wind mechanism. 
While some authors \citep[e.g.][]{1971AZh....48..438S,2001ApJ...560L..71B} have suggested 
that accretion should deactivate this mechanism, others \citep{PSB:2015}
have suggested that the spin-down could be enhanced if the accretion
disk leads to the opening of previously closed magnetic field lines
in the neutron star magnetosphere.

To quantitatively address these questions, we have performed an X-ray timing
campaign on J1023 using {\it XMM-Newton}.  J1023 is the only known
tMSP that is currently in the LMXB state and for which we also have a
precise rotational and orbital ephemeris from the previously observed RMSP state.  In 2014 November/December, we
acquired a pseudo-logarithmically spaced set of four {\it XMM-Newton}
observations, which permitted us to achieve a first phase connection of the X-ray pulsations. 
We further extended this data set with three {\it XMM-Newton} observations taken a year later in 
2015 November/December. Combined with earlier {\it XMM-Newton} observations from 2013 and 2014, 
we have created a simple timing model that apparently phase-connects across the entire LMXB 
state observed from 2013 until now.

In \S\ref{sec:obs} we present these observations and a basic 
data analysis; \S\ref{sec:timing} describes the methodology used to time 
the coherent X-ray pulsations and presents the results so obtained. 
\S\ref{sec:disc} discusses the implications of these results in the context 
of various theoretical models. Lastly, in \S\ref{sec:conc} we provide a 
synopsis of the main results and an outline for future work.\\

\section{Observations \& Basic Analysis}
\label{sec:obs}

\subsection{{\it XMM-Newton} Timing Observations}
\label{subsec:obs}

Though J1023 has previously been observed with {\it XMM-Newton} in the LMXB state \cite[e.g.,][]{BAB:2015}, 
we acquired a new set of shorter observations--presented here for the first time--with a specific cadence to allow an 
unambiguous phase connection of the X-ray pulsations (i.e. an accounting of all neutron star
rotations over the full span of observation).  A
detailed explanation of the observing cadence is given in
\S\ref{subsec:mot}. 

Firstly, J1023 was observed with {\it XMM-Newton} on four occasions at the end
of 2014: November 21 (ObsID 0748390101), November 23 (0748390501),
November 28 (0748390601), and December 17 (0748390701) as a part of
the Director's Discretionary Time (DDT) program. These observations
resulted in 32, 33, 17, and 32\,ks of effective exposure (EE),
respectively, and established a month-long timing baseline. 
The EE here refers to `Good Time
Intervals' (GTIs) during which the telescope was actually collecting
data.  There was no filtering of flares extrinsic to J1023, as it was
deemed unnecessary for the analysis presented here.

A second group of closely spaced observations also became available in 2015, 
where J1023 was observed with {\it XMM-Newton} on November 11 
(ObsID 0770581001, EE: 32.4\,ks), November 13 (0770581101, EE: 24\,ks),
and December 12 (0783330301, EE:27\,ks).  This established
a second month-long baseline of dense observations. 

For all exposures, the European Photon Imaging
Camera's (EPIC) MOS1 and 2 detectors \citep{TAB:2001} were set up in
`Small Window' mode to mitigate the deleterious effect of photon
pileup. The EPIC pn detector \citep{SBD:2001} was configured for the `Fast
Timing' mode, which offers a readout time of 30 $\mu$s by sacrificing
one imaging dimension. During all seven observations, the co-aligned
{\it XMM-Newton} Optical Monitor \citep{MBM:2001} acquired photometric
data in the $B$-band filter in the high-cadence `Image Fast' mode.  All the
cameras were used with the thin optical blocking filter. 

A summary of all existing {\it XMM-Newton} timing mode observations of J1023,
during both its current LMXB state and previous RMSP state, can be
found in Table~\ref{table:obssum}.  For simplicity, we have numbered
these in Table~\ref{table:obssum} and will refer to these as
Obs. $1-10$ for the rest of the paper.  Obs. 1, taken 2008 Nov (ObsID
0560180801), is the only {\it XMM-Newton} timing mode observation
available during the previous RMSP state \citep{AKH:2013}.  The
archival data also includes Obs. 2 and 3, two longer observations
acquired in 2013 Nov (ObsID 0720030101) and 2014 June (ObsID
0742610101), when J1023 was in its current LMXB state
\citep{ABP:2015,BAB:2015}. In this work, we also include these two long
observations to constrain J1023's timing behavior as an LMXB. Thus, 
we have used all the nine \textit{XMM-Newton} observations of J$1023$
(see Table~\ref{table:obssum}) to construct a timing solution in the LMXB state.
 
\subsection{Other Monitoring Observations}
We are currently running a monitoring campaign with the 305\,m
Arecibo radio telescope in Puerto Rico to look for a switch of J1023
back to the RMSP state (or to see whether radio pulsations are
intermittently detected in the LMXB state). With Arecibo, we have
observed J1023 since 2014 July for 71\,hr total to date, with an
integration time of $\sim 0.5 - 1$\,hr per session. A detailed account of the
observational setup will be provided in our upcoming paper. We are observing
the source up to 5\,GHz central frequency. The relatively high observing
frequency is chosen to mitigate the effects of eclipses due to
intra-binary material \citep[see][]{ASR:2009}. We folded each of the
observations using the {\tt dspsr} package \citep{SB:2011} and the
known radio timing ephemeris (see \S\ref{subsec:priortiming} for
details).  A visual inspection of the resulting data cubes shows no
obvious signs of the radio pulsar signal \citep[see][]{SAH:2014}, though we caution that
variations in the orbit mean that the radio ephemeris is not accurate at predicting orbital 
phase and phase shifts of the pulsations as a function of time are expected
(see \S\ref{subsec:pulsesearch}). J1023 is also monitored at lower
observing frequencies with the $76$\,m Lovell Telescope at the Jodrell Bank 
(400\,MHz bandwidth at 1500\,MHz center frequency, at roughly weekly cadence).
  
In parallel, since 2015 October, we have been monitoring the stability of 
J1023's X-ray luminosity using {\it Swift}-XRT target of 
opportunity observations. The 26 roughly $1$\,ks observations
show a relatively stable flux (modulo the high/low modes and flares) at a level expected for the LMXB state. We also 
constructed a light curve\footnote{Constructed using {\it Swift}-XRT data products generator, an online data analysis tool offered by the UK {\it Swift} Science Data Centre.}(binned per observation) for all the {\it Swift} observations 
since J1023 transitioned to an LMXB state in 2013. We see that the count rates
have remained stable over the course of the past 2.5\,years.  In addition to 
the less frequent \textit{XMM-Newton} observations, this further confirms that J1023 has not transitioned 
to an RMSP state since 2013 June, and has remained stable in its X-ray properties.  
 
\subsection{Preparation of {\it XMM-Newton} Data}
\label{subsec:dataprep}
The {\it XMM-Newton} data products presented here were processed using
the Science Analysis Software\footnote{The {\it XMM-Newton} SAS is
  developed and maintained by the European Space Astronomy Centre and
  the Survey Science Centre at the University of Leicester.} (SAS)
version {\tt 20141104$\_$1833-14.0.0}.  The EPIC event lists were
filtered using the recommended {\tt FLAG} and {\tt PATTERN}
ranges. For our analysis we only used photons in the energy range
$0.3-10$\,keV. The X-ray source and background events for each
observation were obtained using the same extraction regions as in
\citet{ABP:2015}. The OM photometric data were extracted using the
{\tt omfchain} processing pipeline in SAS. The times of all
photon event lists and time series light curves were translated
to the solar system barycenter using the DE405 solar system
ephemeris and the VLBI astrometric position of J1023 from
\citet{DAB:2012}.

Figure~\ref{fig:lc} presents the X-ray and
optical light curves obtained in each of the seven  
observations from $2014$ and $2015$. The light curves show great similarity to the
previously observed light curves for J1023 from the longer {\it
XMM-Newton} observations in the preceding 1.5 years of the
LMXB state \citep{ABP:2015,BAB:2015}.  In other words, the same
three luminosity modes are present with approximately the same
duty cycles and at roughly the same luminosities (except in
Obs. $5$ and $6$ where no flares are detected)--suggesting that the 
system's state has remained very stable. Moreover, in
Fig. \ref{fig:lc}, we compare the X-ray and optical
brightness. We note, e.g., that a long string of flares in the
X-ray data also corresponds to observed optical flaring. 

\subsection{Motivation for Chosen Observing Strategy}
\label{subsec:mot}
The four-observation campaign conducted in 2014 was designed in order to
ensure that we could construct a phase-coherent model for the neutron
star's rotation that would be valid for at least the time-span covered
by these four observations themselves (roughly a month).  A
phase-coherent model compares the rotational phase at multiple epochs
and unambiguously counts each individual rotation of the neutron star.
As such, it provides much higher precision than an incoherent timing
approach, i.e. one in which the observed rotational period is compared
between epochs.  This precision is critical for placing a meaningful
constraint on whether the neutron star's spin evolution rate has
appreciably changed between the RMSP and LMXB states.

The chosen strategy employed pseudo-logarithmically spaced
observations, separated by intervals of roughly 2, 5, and 19 days (see
exact dates in \S\ref{subsec:obs}) in order to ensure phase
connection: the idea being that, once phase connection could be
unambiguously achieved between the most closely spaced observations,
subsequently a refined model could be obtained using the larger
spacings. The individual observation durations of $\sim 32$\,ks were
chosen to ensure an accurate spin period determination and high signal-to-noise-ratio (S/N) pulse
profile for each individual observation.  

Another crucial consideration is the fact that J1023's orbital parameters vary
significantly on month-long timescales \citep{ASR:2009,AKH:2013},
meaning that the exact orbital parameters must be determined at each
observing epoch in order to accurately fold the data at the 1.69\,ms
spin period.  This can be achieved with high precision because each
$\sim 32$\,ks {\it XMM-Newton} observation represents close to two
times the 4.8-hr orbital period of J1023, thus removing covariances
when modeling orbital variation via a changing $T_\mathrm{asc}$ at each epoch. Nevertheless, 
we still test for the possibility and significance of such covariances and show them to 
be unimportant (see \S \ref{subsec:pulsesearch}). 

The three short observations in 2015 were spaced by 2 and 26 days (see 
\S\ref{subsec:obs} for details). These observations approximately conform to 
the optimized strategy for the 2014 observation campaign described above and also allow
for a phase connection to be established given prior constraints. Similarly, their average observation time of
$\sim 28$\,ks, is also sufficient to robustly model the varying orbit at each epoch. 

From the instrumental side, there are important limitations and
uncertainties on the accuracy and precision of the EPIC-pn timing
mode.  First of all--due to multiple factors such as spacecraft
clock issues, observation to UTC time conversion, ground station
delays, spacecraft orbital ephemeris, etc.--there is a $\pm
48$\,$\mu$s uncertainty (1 $\sigma$ scatter) on the absolute time
stamp associated with the beginning of each observing session
\citep{MKC:2012}.  This dominates the uncertainties on our pulse phase
measurements.  Second, there is also a clock drift {\it during}
EPIC-pn observations, conservatively determined to be $< 10^{-8}$\,ss$^{-1}$
by \citet{MKC:2012}.  For a 30\,ks observation, a clock drift of 10$^{-8}$ would 
smear the pulse profile by 0.3\,ms, which is 0.18$\times$ J1023's 1.69-\,ms pulse period.
However, the simple fact that we can detect unsmeared pulsations using a constant
pulse period within each observing session suggests that the clock
drifts by $\ll10^{-8}$\,ss$^{-1}$.  Work is ongoing to provide
quantitative limits, and early results corroborate the idea that the
clock drift is unimportant for the work presented here (M.Cruces et al.,
in prep.).

\begin{table*}
\normalsize
\centering
\begin{minipage}{\textwidth}
\caption{\normalsize{Summary {\it XMM-Newton} Timing Mode Observations of J1023}}
\centering
\scalebox{0.85}{
\begin{tabular}{|l|l|l|l|l|l|l|l|l|}
\hline
Obs.& \textit{XMM} Obs Id& \pbox{5cm}{Obs. Date} & \pbox{5cm}{Obs. Start \\ Time (MJD$)^*$} &\pbox{5cm}{Dur.\\(ks)}  &Phase Shift& \pbox{5cm}{$\Delta T_\mathrm{asc}$ \\ (secs)}&\pbox{5cm}{$L_x$\,High Mode\\$(10^{33}$\,erg/s)}&\pbox{5cm}{$L_x$\,Low Mode\\$(10^{33}$\,erg/s)}\\
\cline{1-7}
\hline
\multicolumn{9}{|c|}{RMSP State Observation (presented in \cite{AKB:2010})} \\
\hline
1&0560180801 & 2008-11-26 & 54796.310289 & 34.5&-&-&\multicolumn{2}{|c|}{0.094\,(6)} \\
\hline
\multicolumn{9}{|c|}{AMXP State: Long observations (those presented in \cite{ABP:2015} and \cite{BAB:2015})} \\
\hline
2&0720030101 & 2013-11-10 & 56606.765729 & 138&0.0000$\pm$0.0286 &-28.1$\pm$0.04 & 3.17 $\pm$ 0.02&0.54 $\pm$ 0.01 \\
3&0742610101 & 2014-06-10 & 56818.180417 & 131&0.8913$\pm$0.0285&-25.7$\pm$0.05 & 3.06 $\pm$ 0.02 &  0.45 $\pm$ 0.01\\
\hline
\multicolumn{9}{|c|}{AMXP State: Short observations (those presented here for the first time)} \\
\hline
\multicolumn{9}{|c|}{Short observations $2014$} \\
\hline
4 & 0748390101 & 2014-11-21 & 56982.797081 & 35.7 & 0.4307$\pm$0.0302 & -25.7$\pm$0.24 & 3.30 $\pm$ 0.04 & 0.43 $\pm$ 0.02\\
5 & 0748390501 & 2014-11-23 & 56984.791808 &36.2 & 0.4540$\pm$0.0285 & -25.7$\pm$0.10 & 2.97 $\pm$ 0.03 & 0.49 $\pm$ 0.02\\
6 & 0748390601 & 2014-11-28 & 56989.942650 & 22.0 & 0.4508$\pm$ 0.0285 & -25.9$\pm$0.18 & 3.08 $\pm$ 0.05 & 0.36 $\pm$ 0.02\\
7 & 0748390701 & 2014-12-17 & 57008.683583 & 35.8 & 0.4357$\pm$0.0285 & -25.0$\pm$0.09 & 3.12 $\pm$ 0.03 & 0.41 $\pm$ 0.02\\
\hline
\multicolumn{9}{|c|}{Short observations $2015$} \\
\hline
8 & 0770581001 & 2015-11-11 & 57337.840522 & 32.4 & 0.2224$\pm$0.0288 & -32.6$\pm$0.09 & 3.09$\pm$0.04 & 0.43$\pm$0.03\\
9 & 0770581101 & 2015-11-13 & 57339.169811 & 24.0 & 0.2806$\pm$0.0287 & -32.5$\pm$0.09 & 3.27$\pm$0.04 & 0.34$\pm$0.03\\
10&0783330301 & 2015-12-09 & 57365.070413 & 27.7 & 0.1587$\pm$0.0286 & -31.0$\pm$0.09 & 3.11$\pm$0.03 & 0.45$\pm$0.03\\

\hline
\end{tabular}}
\vspace{0.1 in}
\label{table:obssum}
\end{minipage}
\parbox{0.85\textwidth}{\small{$^*$Start time corresponding to the EPIC-PN in fast timing mode with the THIN1 filter. All the times of photon event lists are barycentered. \\Luminosities quoted here are unabsorbed luminosities based on the known parallax distance for J$1023$ from \cite{DAB:2012}. The luminosities were obtained from spectral fits to the MOS1/2 data as shown in \cite{BAB:2015} and their uncertainties are estimated at a $90$\% confidence level.}}
\vspace{0.25 in}
\end{table*}

\subsection{Prior Timing Information}
\label{subsec:priortiming}
The previously published radio timing of J1023 forms the input
rotational and orbital model for the X-ray timing analysis presented
here, as well as the spin-down model to which it will be compared
\citep{AKH:2013}.  This ephemeris was obtained via long-term timing
observations of J1023 in the RMSP state, conducted for four years using
Arecibo, the Lovell, the Green Bank Telescope (GBT), and the Westerbork
Synthesis Radio Telescope (WSRT).  Table~\ref{table:ephem} presents
this as the fiducial ephemeris used in our analysis. We note that it
differs from the ephemeris present in Table $1$ of
\cite{AKH:2013} because we exclude the orbital period derivative
(since orbital period variations do not appear to be deterministic,
and thus do not extrapolate well) and thus we have converted the ephemeris
so that the epoch of spin frequency determination is the moment of disappearance.

\subsection{X-ray Pulsation Search}
\label{subsec:pulsesearch}

\begin{figure}
\includegraphics[width=\linewidth]{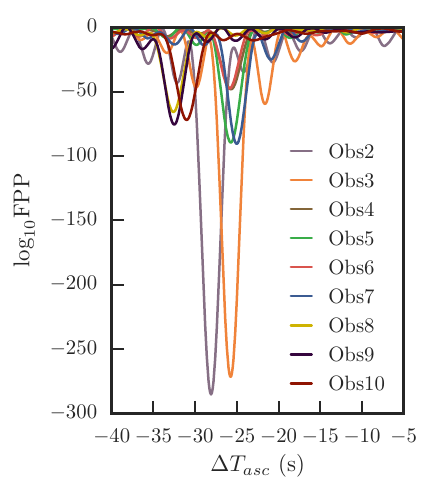}

\caption{The $H$ scores, expressed as false positive probabilities (FPPs), obtained by
  varying $T_{\rm asc}$ with respect to a reference value in each of
  Obs. $2-10$.  The two sets of  closely spaced observations,  Obs. $4-7$ and Obs. $8-10$, 
  show two clusters of dips, indicating that the best-fit $\Delta T_{\rm asc}$ 
  does not vary substantially during these month-long campaigns.  The FPPs are naturally
  smaller for Obs. 2 and 3 because these are $\sim 4\times$ longer than the short observations
  from the two dense campaigns.}

\label{fig:Tasc_dip}
\end{figure}
The aforementioned fiducial ephemeris (see,
\S\ref{subsec:priortiming} and Table~\ref{table:ephem}) was used to assign rotational phases to
the individual X-ray photons from the analysis described in
\S\ref{subsec:dataprep}.  In order to search for X-ray pulsations in
the presence of non-deterministic orbital period variations
\cite[see][]{AKH:2013} we allowed for variation of the fiducial
ephemeris, which resulted in the radio-derived ephemeris, also presented in 
Table \ref{table:ephem}. Since each of our observations spanned at least one orbital 
period, we were able to select the orbital parameters that maximized the 
significance of the detected pulse profile. Although \citet{AKH:2013} 
established that the orbital variations were adequately modeled by 
varying the time of the ascending node ($T_\mathrm{asc}$), we tested the 
effect of varying the projected semi-major axis ($x$) as well. We 
found that this second parameter did show statistically significant 
stochastic variations at the tens-of-microseconds level, but the impact of 
fitting for it on the derived pulse phases was unimportant. 

\begin{figure}
\includegraphics[width=\linewidth]{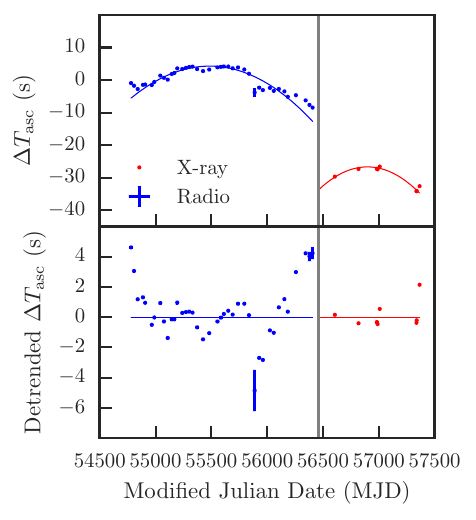}
\caption{Variation in the time of the ascending node, $\Delta T_\mathrm{asc}$,
  with respect to an arbitrary reference point.  {\it Top panel:} raw
  $\Delta T_\mathrm{asc}$ values during both the radio \citep[blue with horizontal lines to denote
  the time over which the orbit was fit;
    see][]{AKH:2013} and X-ray (red) timing epochs.  The gray vertical
  band indicates the state transition from an RMSP to an LMXB state in 2013 June.
  The blue and red parabolas indicate the best-fit parabolas to the radio
  and the X-ray measurements, respectively. We ascribe no direct 
  physical meaning to these orbital period derivatives.  {\it Bottom panel:} $\Delta T_\mathrm{asc}$ values
  after removing the best-fit parabolas to the radio and X-ray measurements.
  This figure is an extended version of that presented in Fig. 1 of
  \citet{ABP:2015}, which included the first two X-ray-derived points
  in the LMXB state.}

\label{fig:Tascall}
\end{figure}

We therefore fit only for variations in $T_\mathrm{asc}$, exploring a range of 
$\pm 40$\,s and selecting the $T_\mathrm{asc}$ that yielded the best $H$ 
score \citep[corresponding to false positive probabilities ranging from 
$10^{-50}$ to $10^{-300}$;][]{JRS:1989}. Fig. \ref{fig:Tasc_dip} shows 
that this allows us to unambiguously determine $T_\mathrm{asc}$, and 
Fig. \ref{fig:Tascall} compares these best-fit $T_\mathrm{asc}$ values to the 
values obtained from radio timing in \citet{AKH:2013}. It also shows two different 
parabolic fits in radio and X-ray states to these best-fit $T_\mathrm{asc}$ values. 

Using the best-fit value for $T_\mathrm{asc}$ in each observation and combining 
them with the radio-derived ephemeris in Table \ref{table:ephem}, we
constructed nine refined local ephemerides for J1023 and successfully detected 
X-ray pulsations at the expected
level, as shown in Fig. \ref{fig:pp}. The individual best $T_\mathrm{asc}$ values for 
all {\it XMM-Newton} observations (including the prior two observations 
in the LMXB state) are listed in Table~\ref{table:obssum}.

\section{Timing Analysis \& Results}
\label{sec:timing}

With X-ray pulsations detected in each of the Obs. $2-10$, we are almost
ready to use these to measure the average evolution of the
neutron star's spin rate during the LMXB state. The next step is to extract pulse
phase information from every observation, as we now describe.

\begin{table*}[h]
\centering

\caption {Ephemerides for J1023}
\begin{tabular}{lrrr}
&  \pbox{5cm}{Fiducial Ephemeris} &  \pbox{5cm}{Radio-derived Ephemeris \\ (Used in this work)}  &  \pbox{5cm}{X-ray Ephemeris} \\
\hline
\hline
Pulsar name    & J1023+0038 & J1023+0038& J1023+0038 \\
MJD range     & 56330.175$-$56458   & 56458$-$57500  & 56458$-$57500   \\
\hline
Spin frequency ($Hz$), $\nu$ & 592.421467941696(11)  & 592.421467941696(11)  & 592.421467941696(11) \\
Spin frequency derivative ($s^{-2}$), $\dot{\nu}$ & $-2.3985\times10^{-15}$ & $-2.3985\times10^{-15}$ & $-3.0413\times{10}^{-15}$ \\
Spin period ($ms$), $P$ & 1.68798744494252 (13) & 1.68798744494252 (13) &1.68798744494252 (13)\\
Spin period derivative ($s/s$), $\dot{P}$  & $6.834\times 10^{-21}$ & $6.834\times 10^{-21}$ & $8.665\times10^{-21}$ \\
Orbital period ($d$), $P_{\rm orb}$ & 0.1980963155& 0.1980963155& 0.1980963155\\
Time of ascending node ($MJD$), $T_{\rm asc}$& 54905.97140075 & * & 54905.97140075\\
Projected semi-major axis ($lt-s$), $x$ & 0.343356& 0.343356& 0.343356\\
\hline
Right ascension, $\alpha$ &10:23:47:687198&10:23:47:687198&10:23:47:687198\\
Declination, $\delta$ &+00:38:40.84551&+00:38:40.84551&+00:38:40.84551\\
Epoch of frequency determination ($MJD$) &56458&56458&56458\\
Epoch of position determination ($MJD$) &54995  &54995  &54995\\
Dispersion measure,  \textit{DM} (cm$^{-3}$pc) &14.325299 &14.325299 &14.325299\\
Solar System ephemeris model & DE 200 & DE 200 & DE 200\\
Orbital eccentricity, \textit{e} &0 &0 &0\\
\hline
\end{tabular}
\\[1cm]

\parbox{14 cm}{\small{Ephemerides for PSR J$1023+0038$. The radio-derived ephemeris is 
    used in \S\ref{sec:timing} to search for pulsations. It is based on the long-term radio timing
    ephemeris (referred here as the fiducial ephemeris) presented in \citet[see][Table~1]{AKH:2013}, but does
    not include an orbital period derivative (since orbital period variations do not appear to be deterministic, 
    and thus do not extrapolate well).  This ephemeris has also been used in \citet{SAH:2014} and \citet{ABP:2015}.  
    Derived as a parameter-restricted fit, it enables an accurate prediction of the
    pulse phase in the neutron star's inertial frame.  In our analysis, we employ this property and construct a 
    radio-derived ephemeris by varying the $T_\mathrm{asc}$ for each one of our observations in order to account for orbital   
    variations (see \S\ref{subsec:pulsesearch}). Hence, we have marked this $T_\mathrm{asc}$ with an asterisk in the radio-derived ephemeris. Apart from the known  
    spin period derivative from the radio state, here we also present X-ray ephemeris, which includes the spin period
    derivative in the LMXB state (see, \S \ref{subsection:PCT} for details). We find that using only this X-ray ephemeris containing the  orbital period derivative, $\dot{P}_\mathrm{orb}$ reported in \S \ref{subsec:Porbdot_fitting}, 
    we can fully account for non-determininistic orbital variations in the LMXB state without fitting for $T_\mathrm{asc}$. In this case,            
    the values for $T_\mathrm{asc}$ and $\dot{P}_\mathrm{orb}$ should be taken as the ones reported in \S
    \ref{subsec:Porbdot_fitting}. 
    Alternatively, we   
    can also exclude the $\dot{P}_\mathrm{orb}$ in X-ray ephemeris and vary the $T_\mathrm{asc}$ (similar to the radio-derived ephemeris) to
    model the orbit in the LMXB state.} }\\

\label{table:ephem}
\end{table*}

\subsection{Computing Per-observation Rotational Phase Shifts}
As a first test, we can simply use the radio-derived ephemeris, with
$T_\mathrm{asc}$ adjusted as described in \S\ref{subsec:pulsesearch}, to fold
the photons from all the nine observations. This folding procedure
uses \texttt{tempo} (\citep{TW:1989}; we also used \texttt{tempo2}, \citep{HEM:2006,EHM:2006}, 
for testing) to compute arrival phases for all photons, with respect to a phase reference. 
After this initial folding, the closely spaced observations 
Obs. $4-7$ were remarkably well aligned, 
showing that there is a very little phase drift over the course of the ${\sim} 1$ month
spanned by these observations.  This already implies that the spin evolution rate of the 
neutron star has not changed significantly with
respect to the radio-derived ephemeris.  In the following subsections, we leverage the full Obs. $2-10$ data set in order to compute
precisely by how much it has changed, taking into account the
impact of the $T_\mathrm{asc}$ fitting procedure on the accuracy of the pulse arrival phases.

\begin{figure}[t]
\includegraphics[width=\linewidth]{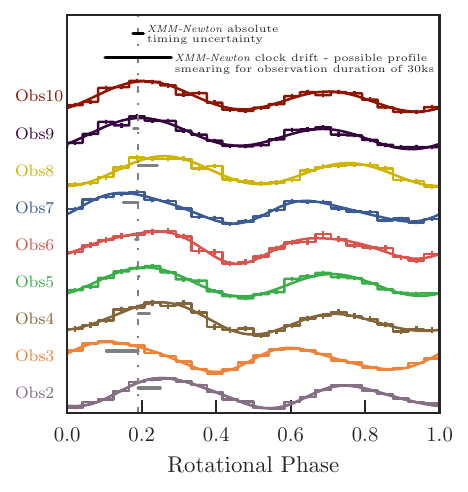}
\caption{Background-subtracted, normalized pulse profiles for each
  of the nine {\it XMM-Newton} observations (from bottom to top:
  Obs. $2-10$, see Table~\ref{table:obssum}) folded using our best-fit spin-down. 
  An arbitrary vertical offset has been added to each profile so that they do not overlap. 
  The color code is the same as in Fig.~\ref{fig:Tasc_dip}. The photons were
  folded using the X-ray ephemeris (see Table~\ref{table:ephem}) containing the spin-frequency derivative 
  in the LMXB state, computed in \S\ref{subsection:PCT}, $\dot{\nu}_\mathrm{LMXB} = (-3.0413\pm0.0090)\times10^{-15}\;\text{Hz~s}^{-1}$. We exclude the orbital period derivative in this ephemeris and instead use the $T_\mathrm{asc}$ computed per observation
  as described in \S\ref{subsec:pulsesearch} to model the orbit in the LMXB state. Histograms are plotted
  for convenience; the smooth curves are obtained directly from photon
  phases \citep[for more details, see][]{ABP:2015} and are used in determining pulse arrival 
  phases. After, the profiles are obtained in such a manner we re-compute their 
  phase shifts w.r.t Obs. 2 (chosen as a template based on its high signal-to-nose-ratio). 
  The horizontal solid gray line shows the offsets of these phase shifts w.r.t. the average phase shift. 
  Note that each profile's phase
  is subject to an $\sim 48\mu$\,s ($\sim$0.028$\times$spin period) timing uncertainty (see \S\ref{subsec:mot}). 
  This absolute timing uncertainty and the clock drift uncertainty are presented here 
  with horizontal solid gray lines at the top of the plot. }
\label{fig:pp}
\end{figure}

Our goal is to compute pulse arrival phases for each of the pulse profiles from 
Obs. $2-10$. These phases are residuals relative to
the radio-derived ephemeris, that is, phase shifts relative to a constant
spin-down equal to that observed in the RMSP state. We do not assume
any particular alignment between the LMXB-state X-ray profile and, the
RMSP-state radio profile or the RMSP-state X-ray profile\footnote{We
  note that, in contrast to the coherent X-ray pulsations discussed
  here, the weak X-ray pulsations detected in the RMSP state are most
  likely not induced by heating of the magnetic polar caps through
  accretion.} presented in \citet{AKB:2010}.

We compute the per observation phase shifts by cross-correlation with
a high-signal-to-noise template. From, the nine observations presented in 
Table \ref{table:obssum}, the long (138\,ks) 2013 November {\it XMM-Newton} 
observation (Obs. $2$) is used as a template given that it has the highest 
signal-to-noise-ratio. We compute the Fourier coefficients of this profile and 
truncate them to the four significant harmonics suggested by the {\it H-test}. 
We then use cross-correlation to optimally align this template with the 
similarly computed Fourier-domain profiles for each of our nine observations. 
The result is a phase offset for each observation.

The uncertainties in these phase shifts come from three main sources:
photon scarcity, covariance with the $T_\mathrm{asc}$ fitting, and (the
dominant effect) the absolute timing uncertainty of {\it XMM-Newton}
(see \S\ref{subsec:mot}). We address the first two with a bootstrap
procedure \citep{boot,Cue:1993}: we repeatedly generate simulated data
sets by drawing photons from the observed photons with
replacement. Then for each simulated data set we repeat the procedure
of fitting for $T_\mathrm{asc}$ and then extract a pulse phase. This
introduces a scatter in pulse phase of roughly ${\sim} 0.0043$ turns. We
then add this in quadrature to the known 48\,$\mu$s absolute timing
uncertainty (equivalent to 0.028 cycles) of {\it XMM-Newton}
\citep{MKC:2012} to obtain the total uncertainty on each pulse phase
measurement. The results are shown in Fig.~\ref{fig:moneyplot} and 
tabulated in Table~\ref{table:obssum}.

\subsection{Phase Connection and Timing}
\label{subsection:PCT}
We now derive a precise measurement of the neutron star's spin-down rate
 during Obs. $2-10$. A change in spin-down rate compared 
to that predicted for the same epoch by the radio-derived ephemeris would 
produce a parabolic trend with time in the measured pulse phases. Therefore, In principle,
we need to simply carry out a least-squares fit to the pulse phases. 
However, since they are phase measurements, the values shown
in the insets of Fig.~\ref{fig:moneyplot} and Table~\ref{table:obssum}
are all recorded modulo one turn of the neutron star.  There may therefore be
an ambiguity coming from the fact that we do not know \emph{a priori} 
how many times the neutron star turned between observations. The spacing of Obs. $4-7$ 
was designed to avoid this ambiguity, but the long gaps between Obs. $2$ and 3 
(due to the \textit{XMM-Newton} Sun constraint) and the non-optimal cadence of Obs. $8 - 10$ allow 
for the possibility of phase wraps when trying to phase connect the full data set.
Therefore, for each candidate spin-down rate, 
we shift the phases appropriately, wrapping them as necessary modulo the spin
period, and compute the sum of squares of the normalized errors ($\chi^2$). 

We followed this procedure first for the dense observation sets from 2014 (Obs. $4-7$)
and 2015 (Obs. $8-10$) individually, obtaining spin frequency measurements from each set. 
We then summed the two $\chi^2$ error functions for these 
observation sets.  The minimum of this summed function yields a change in frequency 
derivative corresponding to a phase connection valid only within the month-long observation sets individually, 
but not necessarily between them. We call this the \textit{Short Baseline} 
approach. Once we could semi-coherently phase connect the two 
clusters of observations, we attempted phase connection between all seven observations in these two 
sets to arrive at $\chi^2$ errors and corresponding minima. This resulted in a 
phase connection of the two observation sets separated by a year (Obs. $4-10$ phase-connected), referred to here as 
the \textit{Year-long Baseline} approach . Finally, in the \textit{Full Baseline Approach}, 
we included all the nine {\it XMM-Newton} observations (Obs. $2-10$) to get a corresponding
$\chi^2$ error function and establish a fully coherent phase connection valid over 
the past $2.5$\,yrs. 

\begin{figure}[t]
\hspace{-0.25cm}
\includegraphics[width=\linewidth]{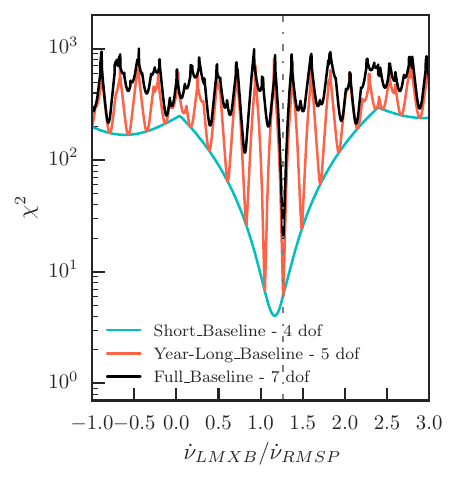}

\caption{$\chi^2$ values for a range of LMXB-state spin frequency derivatives, 
$\dot{\nu}_\mathrm{LMXB}$. Here, the gray dotted-dashed line marks the new value of 
the observed spin frequency derivative obtained from the coincidence of the dips in 
$\chi^2$ error functions for various phase connection baselines. Uncertainty on 
this value differs for each phase connection baseline and is obtained from the 
change in frequency derivative, which increase the minimum $\chi^2$ value by one.  
Finally, we also list degrees of freedom (dof) associated with each phase connection baseline.}
\label{fig:chimin}
\end{figure}

\begin{figure*}
\centering
\begin{minipage}{\textwidth}
\includegraphics[width=\textwidth]{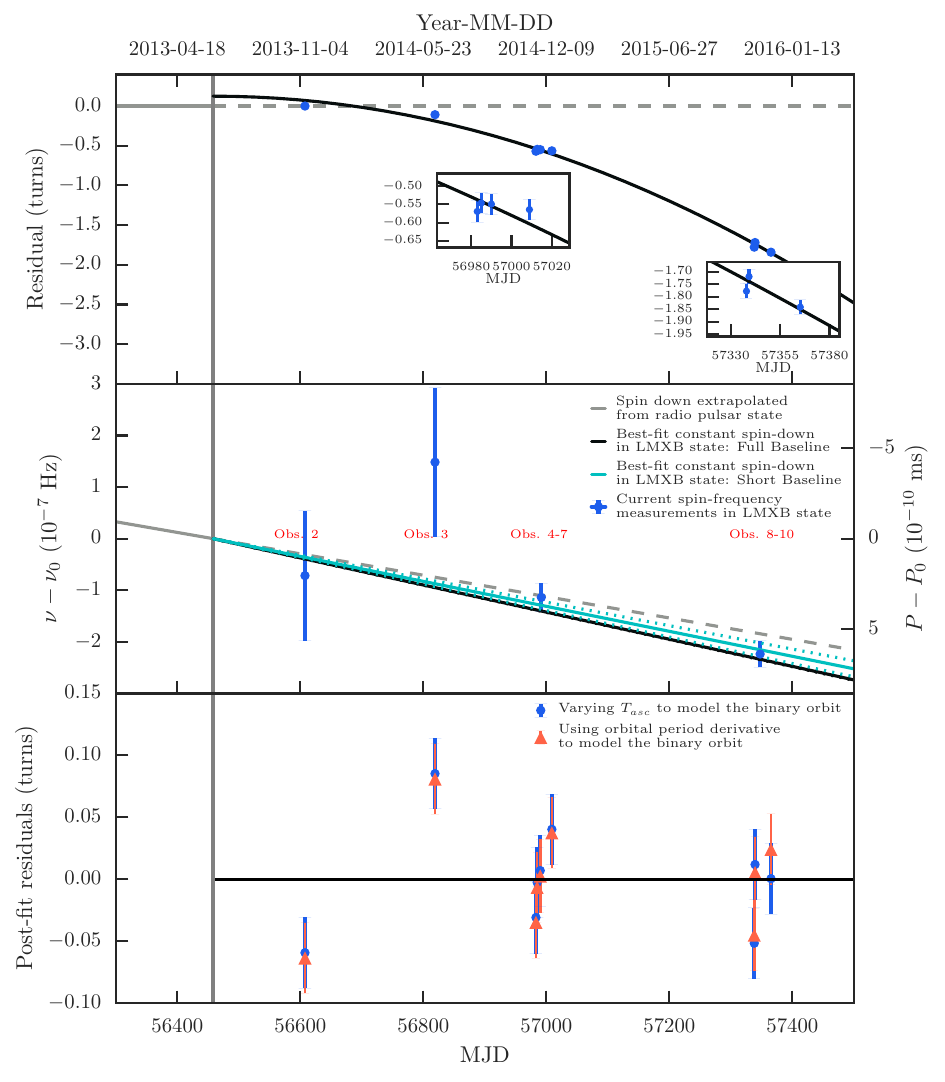}
\vfill

\caption{Our model for the spin evolution of J1023 and the measurements constraining it.
  The vertical, solid gray line in each subplot indicates the
  epoch of RMSP to LMXB state transition (2013 June).  Other solid
  gray lines indicate values derived from the radio ephemeris, dashed
  where these are extrapolated beyond the `MJD range' quoted in
  Table~\ref{table:ephem}.  The black curves indicate a
  post-state-transition, constant spin-down model derived from 
  phase connecting all of observations Obs. $2-10$ ({\it Full Baseline} solution as shown in Fig.~\ref{fig:chimin}); extended dotted black lines 
  show 1$~\sigma$ error bars.  {\it Top panel}: phase-shift measurements from 
  Obs. $2-10$ (blue points) with respect to the radio-derived ephemeris. The
  best-fit constant spindown model is shown by the solid black line.
  The inset shows which observation from Table \ref{table:obssum} each
  point corresponds to. When comparing to the quoted phases in Table 1, note 
  that there have been single phase wraps removed between both Obs. 2/3 and 
  between Obs. 3/4-7.  Additionally, there have been two phase wraps removed 
  between Obs. 4-7/8-10.  {\it Middle panel}: spin frequency evolution.
  The solid line shows the derivative of the best-fit model, while data
  points are short-term frequency measurements. Those for Obs.
  2 and 3 are computed from single observations, assuming the {\it XMM-Newton} clock
  drift is negligible, while those from Obs. $4-7$ and $8-10$ are
  obtained by phase connecting only the observations within each set.
  These short-term model-independent frequency measurements agree with
  the constant spin-down model we fit to the phases. The black and cyan color 
  show constant spin-down models derived from {\it Full Baseline} and {\it Short Baseline} 
  approaches, respectively.  {\it
    Bottom panel}: the blue colour shows the post-fit phase shift measurements obtained from 
    the profiles presented in Fig. \ref{fig:pp}. The post-fit residuals on fitting with an alternative 
    method using orbital period derivative (see, Section \ref{subsec:Porbdot_fitting} for the technique) are shown in red
    colour.
 }
 \label{fig:moneyplot}
\end{minipage}
\end{figure*}

The $\chi^2$ error functions for the above three approaches are shown 
in Fig.~\ref{fig:chimin}. Although there are many local minima, the 
absolute minima for the {\it Short Baseline}, {\it Year-long 
Baseline} and {\it Full Baseline} approaches all match at a value close to 
$\dot{\nu}_\mathrm{LMXB}/\dot{\nu}_\mathrm{RMSP} = 1.27 $ (i.e. 
there is no combination of additional phase wraps that improve upon this solution). 
Although searching over an even wider range of spin-down rates could in principle 
produce an even better fit to the measured phases, greater frequency derivative 
changes than considered would be apparent in our folding of single observations 
and are thus ruled out \citep[even given the claimed relative 
clock drift of {\it XMM-Newton};][]{MKC:2012}. We are therefore able to unambiguously 
phase-connect all the nine \textit{XMM-Newton} observations obtained during
the last $2.5$\,years of J1023's LMXB state. 

Having established the number of phase wraps relative to the radio ephemeris,
we are able to fit a model to the pulsar's spin-down. We choose a model in which
the spin frequency derivative is constant during the LMXB state but is different
from that during the RMSP state. We also assume that the spin frequency is continuous
(that is, it does not jump at the moment of RMSP to LMXB state transition in 2013 June), 
but we do not assume any phase relationship between the LMXB and RMSP states. 
The change in spin frequency derivative we obtain in this constant spin-down model is 
$\dot{\nu}_\mathrm{LMXB}/\dot{\nu}_\mathrm{RMSP} = 1.268 \pm 0.004$, as shown  
in Fig.~\ref{fig:moneyplot}.  J1023's spin-down rate since transition from RMSP to LMXB
is therefore $\dot{\nu}_\mathrm{LMXB} = (-3.0413\pm0.0090)\times10^{-15}\;\text{Hz~s}^{-1}$. 

The reduced $\chi^2$ of this fit is $2.87$ with seven degrees of freedom, which corresponds 
to a false positive probability of $0.005$. 
Finally, we estimate our uncertainties as the range of $\dot{\nu}_\mathrm{LMXB}/\dot{\nu}_\mathrm{RMSP}$ 
values that increase the (non-reduced) $\chi^2$ by no more than one. Using this technique for the 
three phase connection approaches: {\it Short Baseline}, {\it Year-long Baseline} and {\it Full Baseline}, 
we get uncertainties of $7.17$\%, $0.49$\%, and $0.38$\%, respectively. Including the two long observations, 
Obs. 2 and 3, does not introduce a major change in the uncertainty on the frequency derivative estimate 
because these observations are closest to the time of state transition.  Also, though we deem it unlikely, 
if one assumes that only the {\it Short Baseline} approach is robustly phase-connected then J1023 
is still measured to be spinning down faster in the LMXB state at the 3.5-$\sigma$ level.

\subsection{Alternative Orbital Modeling}
\label{subsec:Porbdot_fitting}
Toward the late stages of this work we realised that we could essentially obtain an 
orbital period derivative in the the LMXB state by fitting a parabola to the time of ascending nodes 
for Obs. $2-10$ (reported in Table \ref{table:obssum}). Although we attribute no physical
significance to this value, this orbital period derivative ($-1.65(0.19)\times10^-10$\,day/day, as shown 
by the red parabolic fit to X-ray measurements in Fig.\ref{fig:Tascall}) taken together with the X-ray 
ephemeris (Table \ref{table:ephem}) containing time of ascending node and orbital period 
(Tasc = 54905.96943473\,MJD and Porb = 0.198096646761\,day, respectively) from the same parabolic fit, 
can be used to fold the observations in the LMXB state. The 
profiles, which we obtain from such a folding operation, are almost identical to the ones 
reported in Fig. \ref{fig:pp}. We then compute phase shifts for 
these profiles w.r.t Obs. $2$ (used here as a template given it has the highest S/N). 
These post-fit phase residuals are then plotted with red color in the third panel of Fig. 
\ref{fig:moneyplot}. Here, we see that the post-fit phase shift residuals obtained from 
modeling the non-deterministic orbital variations with X-ray ephemeris containing either (a) variation of $T_\mathrm{asc}$  or  
(b) the orbital period derivative, coincide. This coincidence independently corroborates our 
technique of varying the $T_\mathrm{asc}$ (detailed in \S \ref{subsec:pulsesearch} and the
appendix \ref{appendix:A1_var}) to model the non-deterministic orbits for pulsars in binary
systems such as redbacks and black widows. 

Moreover, we are now using this orbital period derivative estimate to systematically fold the 
gamma-ray photons from the {\it Fermi}-Large Area Telescope, and search for gamma-ray 
pulsations. 

\subsection{Robustness and Potential Pitfalls}
\label{subsection:ROP}
Fig.~\ref{fig:pp} shows that folding Obs. $2-10$ with the X-ray ephemeris (listed in Table~\ref{table:ephem}) without the
$\dot{P}_\mathrm{orb}$ and varying the $T_\mathrm{asc}$ for each observation (as done in the case of the radio-derived ephemeris) yields well aligned pulse profiles. The X-ray ephemeris now contains the spin frequency derivative with the $\dot{\nu}_\mathrm{LMXB}$ value determined in the preceding section. We then see that as such the greatest outlier from such a folding operation is Obs. 3, which is $2.6 \sigma$ off from the prediction of the best-fit constant spin-down model in the LMXB state (Fig.~\ref{fig:moneyplot}; middle panel). Although the total reduced chi-squared for the fit is $2.87$ (with 7 degrees of freedom, corresponding to a false positive probability of $0.005$), which indicates a reasonable fit, the analysis of \citet{MKC:2012} shows a number of outlier points (see their Fig. 8), so the deviation seen in Obs. 3 may also be the result of {\it XMM-Newton} clock limitations.

We also point out that the three frequencies---the spin frequency at disappearance, the spin frequency obtained from observations 4--7 and that obtained from observation 8--10---lie quite exactly on a line, though the last two were computed assuming only short term phase connection within a group of observations.  This supports the idea that the system is spinning down at a constant rate not very different from that seen in the radio state.

The technique of fitting for $T_\mathrm{asc}$ is new; for details, see the Appendix. Although these variations are small, it is necessary to account for them to recover the pulsations. Our bootstrapping process for error estimation accounts for the (small) impact of this fitting on the derived pulse phases.

There are at least two potential pitfalls for the analysis presented
here.  Firstly, {\it XMM-Newton} is known to occasionally experience
one-second clock jumps. If one occurred near the beginning of one of
our observations, without correction in post-processing, then the phase 
computed for that observation would be dramatically incorrect. Such jumps can be detected by adjusting post-processing parameters; we find none in Obs. $2-10$, and the chance of a spurious 
phase alignment caused by such a jump is very low.

\begin{figure}[t]
\hspace{-0.25cm}
\includegraphics[width=\linewidth]{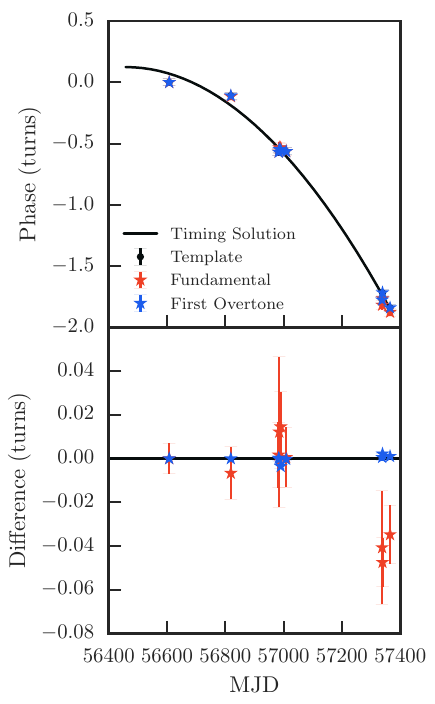}

\caption{Comparison between timing with individual harmonics versus a full template pulse profile. Top panel: the pulse phases obtained by using a full template and the timing solution obtained using them are shown in black. Pulse phases obtained using the fundamental and first overtone are shown in red and blue respectively. Error bars are almost the same since they are dominated by the absolute {\it XMM-Newton} timing uncertainty. Bottom panel: phase differences between the template-derived phases and those obtained from the fundamental and first overtone.}
\label{fig:ccheck}
\end{figure}

Second, we assume that the pulse phase flawlessly tracks the
orientation of the neutron star. This is a standard assumption in the
timing of radio pulsars, but some AMXPs show pulse profile changes,
which indicate that the accretion-induced hotspots are changing size
and/or location \citep{PWK:2009,PAT:2010, PW:2012}.  If the hotspots
we observe in J1023 are wandering, then the apparent spin-down does
not track the neutron star surface. Such hotspot wandering is usually 
accompanied by pulse profile variations and/or luminosity variations. 
We have observed that the X-ray luminosity of J1023 is remarkably constant 
in the low and high modes (particularly during the high mode, which is the only 
one that shows pulsations), and Fig.~\ref{fig:pp} shows that the pulse profiles do 
not vary substantially. We also verified the stability of the 
pulse profiles by using the standard technique of timing harmonics separately 
\citep[see,][where this technique has been used for a sample of six AMXPs]{PWK:2009}.
Our analysis using the fundamental and the first overtone is shown in \ref{fig:ccheck}. 
Here, we see that the pulse phases obtained from using different harmonics are consistent 
with each other and with template-based phases, thus rendering the timing of separate harmonics 
not necessary. This further highlights the stability of the pulse profiles. 
Moreover, it is hard to imagine that such a simple, enhanced constant 
spin-down model could provide such an excellent fit if the hotspot is appreciably 
wandering.

A lingering concern was the incorporation of Obs. 2 and 3 into our phase-coherent timing analysis. Phases determined from these observations are broadly consistent with our best-fit model, although Obs. 3 is somewhat of an outlier, as discussed above. That said, their inclusion makes only a small difference to the best-fit spin-down and its uncertainty, since they are closest to the radio-X-ray state transition and all phase uncertainties are dominated by the {\it XMM-Newton} absolute clock uncertainty.

\section{Discussion}
\label{sec:disc}

Our phase-coherent {\it XMM-Newton} timing observations of J1023
during its current LMXB state indicate that the average spin-down rate
is  $26.8$\% faster than the spin-down rate during the
rotation-powered RMSP 
state--i.e. $\dot{\nu}_\mathrm{LMXB}/\dot{\nu}_\mathrm{RMSP} = 1.268 \pm 0.004$ and $\dot{\nu}_\mathrm{LMXB} = (-3.0413\pm0.0090)\times10^{-15}\;\text{Hz~s}^{-1}$ (see Fig.~\ref{fig:moneyplot}).  
From this we conclude that the previous
spin-down mechanism is still dominant and additional torques on
the neutron star during this accreting state are modest in
comparison. This somewhat surprising result has implications for both
the radio pulsar mechanism and magnetospheric accretion physics. The
X-ray pulsations seen in the `high' mode of the accreting phase are
interpreted as accretion hotspots at the magnetic polar caps
\citep{ABP:2015,BAB:2015}, but the enhanced constant spin-down rate suggests
that the pulsar wind remains active and essentially unchanged even
after the accretion disk forms and matter accretes onto the neutron
star's surface. This is in direct contrast with earlier predictions
(e.g.\ \citealt{1971AZh....48..438S,2001ApJ...560L..71B}) that the
transition to active accretion would suppress the pulsar wind.
However, several authors have proposed that J1023's pulsar has
remained active but is enshrouded by intra-binary material
\citep{CBC:2013,SAB:2013,SAH:2014} in the LMXB state.

\subsection{Stability of the System}
It is again worth emphasizing how stable and predictable the behavior of
J1023 has been over the course of nearly three years. In all the nine {\it XMM-Newton} 
observations in the LMXB state, those presented in \cite{BAB:2015} and those 
presented here, we see basically the same pulse profile morphology and pulsed-fraction.  
Likewise, the luminosity modes (`low', `high') are present in all observations and
flares are detected in all but two short observations (Obs. 5 and 6), at
comparable luminosities. In other words, this is not a chaotic LMXB
outburst, but rather a well-defined, quasi-stable (the X-ray
pulsations switch on-off with the high mode) accretion regime.  At the
same time, though orbital variability is observed, the change in
$T_\mathrm{asc}$ during the LMXB state is no more dramatic than what is seen in the non-accreting RMSP state.

\subsection{Comparison with the Timing of other AMXPs}
Compared with other AMXP systems \citep[see][for a review]{PW:2012},
where pulsations are seen in outburst at much higher X-ray
luminosities\footnote{It is possible that other known AMXPs have
  entered an LMXB state similar to that of J1023 but their much larger
  distance (typically $3-8$\,kpc) has led us to miss this behavior.}
of $10^{35-37}$\,erg\,s$^{-1}$, J1023 has the great advantage of
possessing a known, high precision rotational and orbital ephemeris
derived from radio timing in the RMSP state (as well as a
parallax-derived distance).

Spin variation measurements of other actively accreting AMXPs is
difficult because their typical outburst duration is short ($\lesssim
1$ month).  This makes any timing solution over that limited time-span
insensitive to spin variations of magnitudes comparable to what we have
observed in J1023.  The long-term (i.e. months to years) spin
variation over periods of X-ray quiescence has been securely measured
in only three systems so far \citep{PW:2012}. For example,
SAX~J1808.4$-$3658 is the best-timed AMXP and it has a spin-down on
the order of $10^{-15}\rm\,Hz\,s^{-1}$, measured by comparing the spin
frequency in six different outbursts observed over a baseline of 13
years \citep[e.g.,][]{PBG:2012, HPC:2008, SBR:2008}. Only upper limits
have been placed on the spin-up/down \textit{during} an outburst.  The
other two AMXPs with a measured long-term spin variation also 
show a spin-down in quiescence that is close in magnitude to that
observed in J1023 \citep[however, they both show a moderately strong
  spin-up in outburst at luminosities of
  ${\sim}10^{36}\rm\,erg\,s^{-1}$;][]{PAT:2010, PRB:2011, HGC:2011,
  RBD:2011}.

Whether the long-term spin-down observed in other AMXPs has the same
physical origin as in J1023 is difficult to say at the moment, since
none of these sources have been detected pulsating in radio so far and
thus it is impossible to compare the observed (quiescent) long-term
spin-down with a radio-derived ephemeris. The lack of observed radio
and/or X-ray pulsations during quiescence in AMXPs may simply be due
to their larger distances (typically $3-8$\,kpc, compared to to the
1.3\,kpc of J1023).

\subsection{The Nature of J1023's LMXB state}
J1023 spins down 26.8\% faster in the LMXB state than in the RMSP state. 
This is measured with high precision ($0.4$\% uncertainty) because we can phase connect all the available {\it XMM-Newton} observations. The corresponding spin-down rate 
$\dot{\nu}_\mathrm{LMXB} = (-3.0413\pm0.0090)\times10^{-15}\;\text{Hz~s}^{-1}$. The lack of drastic (factor of a few or more) change in spin-down between states strongly
suggests that the main radio pulsar spin-down mechanism (i.e. the
pulsar wind) remains active in the accreting LMXB state, and that interaction between the accretion flow and magnetic
field introduces a net additional spin-down of $\dot{\nu} \simeq
~-6.4\times10^{-16}~{\rm~Hz~s^{-1}}$.
This observation introduces both new questions and
unprecedented constraints on accretion-regulated spin change.

First of all, it is surprising that the change in magnetic field
configuration needed to support accretion onto the neutron star's
surface is not accompanied by a more dramatic change in spin-down
rate. Radio pulsar spin-down is chiefly determined by an outflow of
relativistic particles along open magnetic field lines
\citep{1969ApJ...157..869G}. Simulations of pulsar winds show that the
rotating magnetic field assumes a largely open, `split monopole'
configuration, with the wind strongest along the plane of rotation
\citep[e.g.][]{2006ApJ...648L..51S}. Simulations and analytical
arguments suggest that the strength of the pulsar wind (and hence the
spin-down rate) is directly proportional to the number of opened field
lines, so that it scales with the area of open field lines
\citep{2006ApJ...648L..51S,PSB:2015}.

In the absence of an accretion disk, the magnetic field is forced open
by relativistic constraints, and all the magnetic field lines that
intersect the light cylinder ($r_{\rm lc} \equiv c/\Omega_* =
8\times10^6~{\rm~cm}$ for J1023) will be opened and support an
outflowing wind. However, the presence of an accretion disk will
radically increase the number of open field lines, changing the
spin-down rate and wind strength. This is because the inner edge of the
accretion disk will interact with the magnetic field, and the large
difference in angular velocity between the star and the disk material will tend to cause
field lines to stretch and become open
\citep{1990A&A...227..473A,1995MNRAS.275..244L,2002ApJ...565.1191U,PSB:2015}. A
small portion of these field lines may then periodically reconnect to
the disk, so that the reconnection and field line opening can lead to
outflows of magnetic field and matter (e.g.
\citealt{1996ApJ...468L..37H,1997ApJ...489..199G,1997ApJ...489..890M,2004ApJ...616L.151R,2013A&A...550A..99Z,2014MNRAS.441...86L};
see also the review from \citealt{2004Ap&SS.292..573U}).  As a result
of the disk-field coupling, virtually all field lines that would
intersect the disk will be opened and remain open, substantially
increasing the spin-down rate. If the X-ray pulsations in J1023 indeed
originate from accretion onto the neutron star, the accretion disk
must at some times (i.e. during the `high' luminosity mode) extend at
least to the co-rotation radius ($r_{\rm c} \equiv
(GM_*/\Omega_*^2)^{1/3} = 2.4\times10^6~{\rm~cm}$) where the star's
spin rate equals the Keplerian disk velocity. This is roughly a third of
the light cylinder radius, so that the area of the
open field line region (and hence the spin-down power of the wind)
should increase by up to 10 times, which is not observed.

How can this discrepancy be resolved? One speculative possibility is
that the high/low modes are not the result of accretion and in fact
represent an exotic example of `mode switching'--a poorly-understood
process observed in some isolated pulsars. We discuss this further in
\S\ref{sec:moding}. A second possibility is that the pulsar wind
spin-down efficiency decreases during accretion episodes due to a
decrease in the plasma supply to the pulsar wind. This has been
suggested by \cite{2012ApJ...746L..24L} to explain variations in the
spin-down rate of isolated pulsars. In this case, the extra spin-down
from the increased open field line region would be offset by a
reduction in spin-down from outflowing plasma. However, given that
there is only a 26.8\% change in spin-down rate whereas the field line
opening would increase spin down by up to 10 times, it seems
somewhat contrived that these two effects nearly perfectly balance each
other.

We would also like to suggest that maybe a residual disk remains present even in the
RMSP state so that the field configuration does not
substantially change between the LMXB and RMSP state. This could happen if the interaction between the magnetic field
and accretion disk allows a `trapped disk' to form
(\citealt{1977PAZh....3..262S,2010MNRAS.406.1208D,DS:2011,2012MNRAS.420..416D},
and also below). In this model, the angular momentum added by the
magnetic field at the inner edge of the disk matches the rate at which
turbulence transports angular momentum outward, so that the net
accretion rate through the disk is zero \citep[this is the `dead disk'
density solution, which was first derived by][]
{1977PAZh....3..262S}. The disk could persist even when the
radio pulsar is active, since the majority of the disk would be
disconnected from the magnetic field and shielded from the pulsar wind
by the magnetic field lines that connected to the inner edge of the
disk. This solution also has observational predictions for the limits
on spin change, which we discuss in \S\ref{subsec:AISM}. 

A final consideration is the nature of the outflow from J1023.  As shown 
in \citep{DMM:2015}, radio observations of J1023 in the LMXB state reveal 
a variable flat-spectrum emission strongly suggestive of a jet and, 
considerably brighter than predicted from observations of other neutron star 
LMXBs accreting at higher accretion rates. Taken at face value, this implies 
that the accretion flow in J1023 is generating a more powerful jet than expected. 
While a propeller mechanism could explain the radio observations, as shown above,
it should lead to a strong spin evolution, which our observations rule out. 
Thus, the interaction of the accretion flow with the radio pulsar in the J1023 
system must both lead to a radio-bright outflow while simultaneously not greatly 
affecting the overall spin-down rate.

\subsection{Limits on Accretion-induced Spin Modulation}
\label{subsec:AISM}
The most natural interpretation for the observed additional
  spin-down is that it is a result of interaction of the magnetosphere with 
 the
  accretion flow and the disk. Such a strong limit on accretion
  torques has never before been set in a millisecond pulsar,
  particularly not at such low luminosities. In this interpretation,
  the net accretion-related spindown
  $\dot{\nu} = -6.4\times10^{-16}\rm{Hz~s}^{-1}$. Using the X-ray
luminosity as a proxy for the accretion rate onto the star, the
well-constrained magnetic field and rotation rate allow us to estimate
the expected spin change from magnetospheric accretion.

The observed luminosity is dominated by channeled accretion onto
  the surface (at a rate $\dot{M}_{\rm obs}$), which spins the star
  up. This means that the spin-down rate is somewhat higher than is
  observed. To correct for this and convert the observed luminosity
  into an accretion rate requires some assumptions for the bolometric
  correction
  factor.
\citet{BAB:2015} measured average luminosities ($0.3-10$\,keV)
of $[0.54,3,10]\times10^{33}~{\rm erg~s}^{-1}$ for the low, high and
flaring states, and estimated that the source spends a [0.22,0.77,0.01]
fraction of its time in each state, giving an average luminosity of
$L_{\rm x} = 2.9\times10^{33}~{\rm erg~s}^{-1}$. {\it NuSTAR}
observation of J1023 showed an unbroken power law in X-ray emission
extending to at least 79\,keV \citep{TYK:2014}, which implies that the
bolometric luminosity is at least
$L_{\rm tot}\geq 6\times10^{33}~{\rm erg~s}^{-1}$. This gives a
minimum accretion rate onto the star of
$\dot{M}{\sim} 3\times10^{13}~{\rm g~s}^{\,-1}$, and a spin-up rate
($\dot{\nu} {\sim} \dot{M}(GM_*r_{\rm c})^{1/2}/(2\pi I_*)$, where $I_*$
is the star's moment of inertia) corresponding to
\begin{equation}
\begin{split}
  \label{eq:spinup1}
\dot{\nu} {\sim} 10^{-16}~\frac{\dot{M}}{3\times10^{13}~{\rm g~s}^{-1}} \left(\frac{M_*}{1.4M_\odot}\right)^{1/2}\left(\frac{r_{\rm c}}{2.4\times10^6~{\rm cm}}\right)^{1/2}\\
\left(\frac{I_*}{10^{45} {\rm g~cm^2}}\right)^{-1} {\rm ~Hz~s^{-1}}
\end{split}
\end{equation}
This implies that the total accretion-related angular momentum loss is $\sim -7\times10^{-16} {\rm ~Hz~s^{-1}}$.


J1023's low luminosity of the source (and inferred low accretion rate)
suggests that it is accreting in what is usually called the
`propeller' regime \citep{IS:1975}. This situation can occur when the
magnetic field truncates the inner disk well outside the co-rotation
radius, so that the star's magnetic field spins much faster than the
inner edge of the accretion flow, creating a centrifugal barrier that
prevents accretion. As long as the inner edge of the disk
  (called $r_{\rm m}$, the magnetospheric radius) is far from the
  co-rotation radius, the relative velocity between the disk and the
  magnetic field is large, and the magnetospheric radius can be
  estimated \citep{1976ApJ...207..914A}:
\begin{equation}
  r_{\rm m} = \xi\left(\frac{\mu^4}{2~GM_*\dot{M}^2}\right)^{1/7}
\end{equation}
where $\xi \simeq 0.4$--1 is a correction factor for disk accretion
(see e.g. \citealt{1979ApJ...232..259G}).

However, when $r_{\rm m}$ is similar (within a factor two) to
  $r_{\rm c}$, the relative rotation between the star and the inner
  disk is not so large, and a correct estimate for $r_{\rm m}$ must
  consider the relative velocity between the two
  \citep{ST:1993,1996ApJ...465L.111W}. This estimate
  defines $r_{\rm m}$ as the point at which the magnetic field is
  strong enough to enforce gas co-rotation with the star. This yields
a somewhat smaller estimate for $r_{\rm m}$ than the one given above:
\begin{equation}
\label{eq:rm_mod}
  r_{\rm m} \simeq \left(\frac{\eta \mu^2}{4\Omega_*\dot{M}}\right)^{1/5}
\end{equation}
where $\eta < 1$ is the relative size of the $B_\phi$ component
induced by the relative rotation between the disk and the magnetic
field. Given that J1023 is accreting and being spun down, this
suggests that $r_{\rm m} \sim r_{\rm c}$, thus eq.~\ref{eq:rm_mod}
is more appropriate to estimate the location of $r_{\rm m}$.

We can use the inferred magnetospheric spin-down
  ($\sim -7\times 10^{-16}~\rm{~Hz~s^{-1}}$) to set a limit on the mass
    outflow rate (assuming that mass ejection is responsible for
    spin-down). For gas outflowing at the escape velocity, the
    spin-down rate will be
    $\dot{\nu}\sim -\dot{M}_{\rm out}(GM_*r_{\rm m})^{1/2}/(2\pi
    I)$. The outflow rate required to spin down the star by the
    observed amount is then
\begin{equation}
\begin{split}
\dot{M} \simeq 1.4-1.8\times10^{14}~{\rm g~s^{-1}}\left(\frac{I}{10^{45}\rm{g~cm^2}}\right)^{10/9}\\\left(\frac{\dot{\nu}}{7\times10^{-16}{\rm Hz~s^{-1}}}\right)^{10/9} 
\left(\frac{R_*}{10{\rm km}}\right)^{-2/3}\\
\left(\frac{B}{10^8{\rm G}}\right)^{-2/9}\left(\frac{P_*}{1.7{\rm ms}}\right)^{-1/9}
\end{split}
\end{equation}

The range in $\dot{M}$ comes from assuming $\eta = 0.1-1$,
corresponding to a weak or strong coupling between the disk and
magnetic field, and would require (on average) that only about
one-fourth--one-fifth of the gas in the disk is accreted onto the star. These
accretion rates correspond to an uncertainty in $r_{\rm m} \simeq$
1.4--2.3$r_{\rm c}$, suggesting that the centrifugal acceleration
needed to launch an outflow is only likely to be strong enough if the
coupling between the disk and the star is strong. (Using the
conventional formula for $r_{\rm m}$ gives much weaker constraints on
$\dot{M}$ and $r_{\rm m}$, and predicts outflows of up to
$10^{12}{\rm g~s^{-1}}$.) Numerical simulations and analytic work do
not currently give strong constraints on expected outflow rates, but
based on our results we conclude that the star-disk coupling may be
able to drive a strong outflow, but only if the coupling is strong. If
$r_{\rm m}$ stays around 1.4$r_{\rm c}$, the coupling is unlikely to
be energetic enough to launch most of the gas into an outflow, and it
can remain bound, creating a trapped disk.

Whether a trapped disk forms depends on using angular
  momentum conservation to predict the response of the disk to
  interaction with the magnetic field. When the accretion rate
decreases such that $r_{\rm m} > r_{\rm c}$, angular momentum
conservation predicts a very different outcome
from the propeller. If the rotation rate between the inner disk edge
and the magnetic field is similar, the energy added by the magnetic
field will not be enough to drive an efficient outflow
\citep{ST:1993}, which can lead to an accumulation of gas in the inner
regions of the disk. This will alter the gas density profile of the
inner disk regions to a dead disk solution
\citep{1977PAZh....3..262S}. In a dead disk, $r_{\rm m}$ is no longer
set by the mass accretion rate, but instead is the distance from the
star where the rate angular momentum added to the disk by the
disk-field interaction is balanced by the rate that viscous turbulence
can transport it outward.

This solution was studied in detail by
\cite{2010MNRAS.406.1208D,DS:2011,2012MNRAS.420..416D},
who found that in this scenario the inner edge of the disk becomes
trapped near $r_{\rm c}$ even when the accretion rate decreases by
orders of magnitude (hence the term `trapped disk'). For J1023, this
implies that the observed accretion rate closely matches the actual
one, and a large outflow of gas is not necessary. The angular momentum
lost by the star is then added to the accretion disk, at a rate given
by
\begin{equation}
  \dot{\nu} \simeq \frac{\eta\mu^2\Delta r}{2\pi I_*r_{\rm c}^3}
\end{equation}
where $\Delta r < 0.3$ is the width of the inner disk region that
remains coupled to the magnetic field, and $\eta$ is the strength of
the coupling (introduced in eq. \ref{eq:rm_mod}). Unfortunately, this
equation has two unknown parameters: $\eta$ and $\Delta r$, which
could both be in the range ${\sim}0.01-1$ depending on the geometry
of the system (e.g. the relative inclination of the magnetic field with
the disk) and the poorly constrained details of the disk-field
interaction. The maximum predicted net spin-down rate for J1023 is
then $\dot{\nu} {\sim} 2\times 10^{-14}$, while for more typical
values assumed by \cite{2010MNRAS.406.1208D}
($\Delta r/r \sim 0.01-0.2$, $\eta \sim 0.1$), the predicted spin
  down matches the observed one quite well, with
$\dot{\nu} {\sim} 10^{-16}-10^{-15} {\rm Hz~s}^{-1}$. A drawback to
the trapped disk scenario is that it does not immediately offer an
explanation for why the outflow from J1023 should be more radio-bright
than expected based on higher-accretion rate neutron star LMXBs. The
moding behavior of J1023 may offer a clue in this regard--if
unstable accretion from the trapped disk is intermittently driving a
propeller (for example, in the low mode), then the additional spin-down
it induces will be reduced by the moding duty cycle. However, if a
trapped disk remains even in the RMSP state, any spin-down from the
disk-field interaction would be folded into the overall spin-down
rate, so no change would be expected. At the moment there are no
observable predictions for what a trapped disk in the RMSP state would
look like, but this possibility warrants further investigation.

\subsection{Mode Switching}
\label{sec:moding}
A more exotic and speculative suggestion is that small amounts
of accreting material could be stimulating the pulsar magnetosphere to
switch between two stable luminosity modes (corresponding possibly to
two stable geometric configurations of the magnetic field structure).
In this model, the infalling material is not being accreted onto the
surface of the neutron star and therefore this is not what is causing
the observed X-ray pulsations.

Mode switching is a long-known (but poorly understood) phenomenon in
rotation-powered pulsars, in which the pulse profile switches between
two stable and reproducible morphologies \citep[e.g.,][]{BHK:2014}.
It is sometimes accompanied by noticeable changes in the spin-down rate
\citep{KOJ:2006,LHK:2010}, though not always. In some moding pulsars only 
upper limits on the spin-down change are available \citep[cf.][]{YSW:2012}. 

Recent observations of PSR~B0943+10 showed that the radio pulse profile 
changes can be accompanied by simultaneous switches in the X-ray pulse profile and
brightness \citep{HHK:2013}.  Similarly, the radio-quiet $\gamma$-ray
pulsar PSR~J2021$+$4026, in the Gamma Cygni region, has also shown a
sudden profile change accompanied by a spin-down rate change
\citep{FLC:2013}.  Moding has never been seen in an RMSP or AMXP, nor is it
believed to be associated with accretion.  Nonetheless, the idea that
moding, nulling, and intermittency may be associated with changes in
the plasma density of the pulsar magnetosphere lead us to speculate
that this effect could also be induced by small amounts of accretion
matter entering the magnetosphere.  That said, in this scenario, it is
completely unclear how material parked at the light cylinder can move
in and out about this radius to trigger the mode switching.
Nonetheless, the rapid switches in J1023's X-ray light curve between
low and high modes is reminiscent of the abrupt switches seen in
pulsar moding, and we also know that the X-ray pulsations are also
switching on/off at these times.  However, this is only a qualitative
similarity. Though it is likely impossible from a practical
point-of-view (there are of order 10 {\it Fermi}-detected $\gamma$-ray photons from J1023
during all the {\it XMM-Newton} X-ray observations presented here), if one could show that the
$\gamma$-rays are generated
only in the low or high mode, then one could plausibly argue that the
X-ray pulsations are due to heating of the magnetic polar caps, not by
the accretion flow itself but by the normal rotation-powered mechanism
also responsible for the observed X-ray pulsations in the RMSP state.

\section{Conclusions \& Future Work}
\label{sec:conc}

By phase connecting a set of nine {\it XMM-Newton}
observations, we have shown that J1023 spins down on average
26.8\% faster since 2013 June, when it transitioned from a rotation-powered 
RMSP state to an accretion-powered LMXB state.
This is the first time that it has been possible to make such a 
comparison for a tMSP system, and it shows that the accretion torques 
during the LMXB state are modest compared to the primary spin-down 
driver that is also present in the RMSP state (i.e. the pulsar wind).  
The rich observational phenomena displayed by J1023 has also been 
seen in the other confirmed tMSPs, namely \xss~and \igr.  
As such, there is good reason to expect this behavior to be
typical of tMSPs in the `intermediate' accretion LMXB state
\citep{Lin:2014}.

J1023 can continue to be timed in the LMXB state, in order to confirm
that the average spin-down rate observed up until now will persist.
Unfortunately, the only instrument capable of providing the necessary
observations is {\it XMM-Newton} in the EPIC-pn timing mode.  Such
observations are only possible in two ${\sim} 1.5$ month windows per
year, when J1023 (a nearly ecliptic source) is not in the Sun constraint.
Continued timing is most strongly motivated by the desire to provide a
continued baseline that can be connected to the timing measurements
that will undoubtedly follow once the source reignites as an
observable RMSP. If continued timing across a future LMXB to RMSP state transition
shows a sudden jump in the spin frequency and its derivative, it can potentially 
help us understand the torques acting on the neutron star during such a transition. 
Lastly, though the stochastic orbital variations of J1023 will make it challenging, an unbroken timing
solution is necessary for attempts to detect pulsed $\gamma$-rays from
the ongoing {\it Fermi} observations.

\section{acknowledgements}
The authors gratefully thank Lucien Kuiper for his insights into the
EPIC-pn timing mode as well as Ben Stappers and Andrew Lyne for their previous
contributions to timing J1023 in the RMSP state. A.J., J.W.T.H. and C.B. acknowledge funding from the
European Research Council under the European Union's Seventh Framework
Programme (FP7/2007-2013) / ERC grant agreement no. 337062
(DRAGNET). A.M.A acknowledges support from an NWO Veni Fellowship.
A.P. and C.D'A. acknowledge support from an NWO Vidi Fellowship. This
work was funded in part by NASA grant NNX15AJ50G awarded through
Columbia University.  The results presented in this paper
were based on observations obtained with \textit{XMM-Newton}, an ESA
science mission with instruments and contributions directly funded by
ESA Member States and NASA. We thank Norbert Schartel and the {\it
XMM-Newton} observatory for granting these DDT observations on short
notice. The Arecibo Observatory is operated by SRI International under a  
cooperative agreement with the NSF (AST-1100968), and
in alliance with Ana G. M\'endez-Universidad Metropolitana,  
and the Universities Space Research Association. 
We have made extensive use of the NASA  Astrophysics Data System (ADS) and the arXiv.  
A.J. and J.W.T.H would like to extend thanks to Joel Weisberg
for helpful discussions. A.J. would also like to thank F. Coti Zelati for
comments on an initial draft. Lastly, we acknowledge the International
Space Science Institute (ISSI) that funded an international team
devoted to the study of transitional millisecond pulsars where this
work has been discussed, and we thank all the members of the team for 
fruitful discussions.

\bibliographystyle{apj}
\bibliography{ApJ_ADJ}
\appendix

\section{Separation of Orbital and Spin-down Effects}
\label{appendix:A1_var}

Traditionally, in pulsar timing, a model is constructed to predict pulse arrival phase, including astrometric, rotational, and orbital effects. This works well for some pulsars, but for pulsars with more complex behavior--e.g, stochastic ``red'' variations of the pulsar spin-down or stochastic variations of orbital parameters--attempting to parametrize that complex behavior can become complex and awkward. It can also become difficult to distinguish variations in orbital parameters from variations in intrinsic pulsar spin-down. We therefore describe an approach that allows clean separation of orbital effects from intrinsic pulsar spin-down. This proceeds essentially by removing the propagation delay across the orbit (where the orbital model may vary from epoch to epoch), yielding pulse emission times as measured at the pulsar position; these emission times can then be treated as if the pulsar were isolated.

Pulsar orbital models vary in complexity. Some pulsars are adequately modeled with Keplerian orbital models, but many require more complex models. Many pulsar systems are sufficiently relativistic to require ``parameterized post-Keplerian'' models in which the Keplerian orbital elements change with time. One millisecond pulsar is known in a stellar triple system \citep{RSA:2014}, where even Newtonian interactions between the two orbits produce substantial deviations from Keplerian models. Other systems, particularly the black widow and redback interacting-binary systems, undergo stochastic orbital variations due to poorly understood processes \citep[possibly quadrupole-moment changes within the companion; for a discussion see][]{AKH:2013}, requiring orbital models flexible enough to accommodate substantial deviations from a Keplerian model (traditionally parameterized by allowing multiple derivatives of orbital parameters).

Given a model for the orbit of a pulsar, observed pulse arrival times are converted to a time scale called pulsar emission time. This conversion involves the removal of delays due to propagation across the pulsar system (both geometric and relativistic Shapiro delay), propagation across the solar system (again, both geometric and Shapiro delays), interstellar propagation (dispersion), and relativistic time dilation at both the Earth and the pulsar. It should be noted that the pulsar emission time scale is conventionally rescaled so that its mean rate equals that of the Earth's, so only the varying component of the time dilation appears in the conversion. Once arrival times have been converted to this pulsar emission time scale, the pulsar spin-down model allows conversion of emission time to phase. This typically uses simple linear model parameterized by spin frequency at some specified epoch and spin frequency derivative. Pulsars with more complex spin-frequency behavior may require more complex models: stable young pulsars may have a measurable second frequency derivative (permitting the calculation of a braking index); glitching pulsars may require a model with phase, frequency, and/or frequency-derivative jumps; pulsars with spin-down noise require a more complex model, often represented as a polynomial parameterized by multiple higher-order frequency derivatives.

The pulsar emission time provides a place to separate the problem of describing the pulsar orbit from that of describing the pulsar's intrinsic spin-down. Specifically, we wish to work with a system in which the orbital parameters vary stochastically. We seek to study the intrinsic spin-down of the pulsar. In this paper, our observations are {\it XMM-Newton} X-ray observations each spanning more than a complete binary orbit. For each epoch, then, we fit for the varying orbital parameters to obtain an orbital model specific to that epoch, using the pulse profile signal-to-noise (H score) as a measure of goodness-of-fit. Thus, in principle, we should be able to use these orbital models to obtain pulsar emission times for each observed pulse.

Existing pulsar-timing tools (e.g. {\tt tempo} and {\tt tempo2}) are not designed to work with pulsar emission times, though they necessarily compute them internally. We are therefore forced to use a more complex analysis procedure. We begin with an ephemeris based on the timing of the system in the RMSP state. This allows us to compute phases for each photon from an X-ray observation; since previous work showed that adjusting the time of the ascending node ($T_\mathrm{asc}$) was sufficient to model the orbital variations, we adjust $T_\mathrm{asc}$ for each observation to maximize the $H$ score of the pulsations. We use this $T_\mathrm{asc}$ adjusted ephemeris to compute photon phases and obtain a folded pulse profile. We then cross-correlate this folded profile against a template (obtained from the observation with the best signal-to-noise) and obtain a phase shift of the observed pulse relative to the pulse predicted by the ephemeris. This requires some care: since {\tt tempo} is not ordinarily concerned with absolute pulse phases, it always sets the phase of the first event to zero. We therefore introduce a synthetic event, the same for each observation, to define the zero of phase (we choose MJD 56480, the nominal date of radio disappearance). 

Here, a subtlety arises: the zero of phase should be defined in the pulsar emission time scale in order to be truly consistent between observations with different orbital parameters, but neither \texttt{tempo} nor \texttt{tempo2} allow the specification of times in this time scale. We therefore perform an inversion process so that our synthetic event occurs at MJD 56480 in the pulsar emission time scale. A second concern is that uncertainties in orbital parameter determination affect the computed pulsar emission times and therefore the computed phases; we therefore include the process of fitting for $T_\mathrm{asc}$ in our bootstrap-based estimation of the uncertainties on the pulse phase. Through this process we obtain a set of phase residuals relative to the given spin ephemeris, which allows us to compute needed changes in spin frequency or frequency derivative as if the pulsar were isolated. We confirm that these changes produce the intended effect by reprocessing the photons with the modified ephemeris; indeed, the residuals shift by the claimed amount and yield a reasonable fit.

As a cross-check of this method, we note that the orbital evolution post-disappearance can be modeled by a single \texttt{tempo} parameter file that includes appropriate $T_\mathrm{asc}$, $P_\mathrm{orb}$ and $\dot P_\mathrm{orb}$. These values can be found by fitting the $T_\mathrm{asc}$ values from individual observations, and this ephemeris yields residual $\Delta T_\mathrm{asc}$ values on the order of a second. Folding all the photons with this single ephemeris results in some amount of profile smearing, but the pulsations remain detectable, and phases (pulse arrival times) can be extracted from these profiles. We are therefore able to directly compare the pulse phases computed using a traditional \texttt{tempo} parameter file to those obtained using our new method; the last panel of Fig.~\ref{fig:moneyplot} shows that the phases are very similar, validating this new technique.

We believe that working with the pulsar emission time scale is helpful in dealing with pulsars with complex spin down behavior. We suggest a relatively minor change to \texttt{tempo} and \texttt{tempo2} to make this easier: in addition to the ``observatory codes'' `\texttt{@}' and `\texttt{0}', which indicate that times are measured at the solar system and Earth barycenters, respectively, we suggest an additional ``observatory code,'' e.g. `\texttt{*}', to indicate that a time is measured in pulsar emission time. We note that \texttt{tempo2}'s `\texttt{general2}' output plugin already permits the output of pulsar emission time for any event (although we caution readers that older versions contained a sign error in its calculation). The addition of the observatory code `\texttt{*}' \texttt{tempo}/\texttt{tempo2} would greatly ease working with timing data from redbacks, black widows, and binary pulsars with complex spin-down behavior.

\begin{figure*}
\centering
\begin{minipage}{\textwidth}
\includegraphics[width=\textwidth]{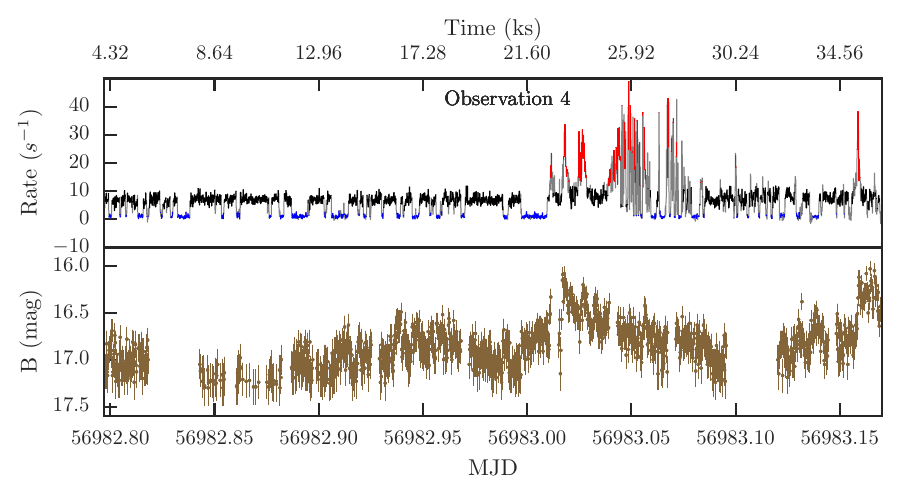}
\includegraphics[width=\textwidth]{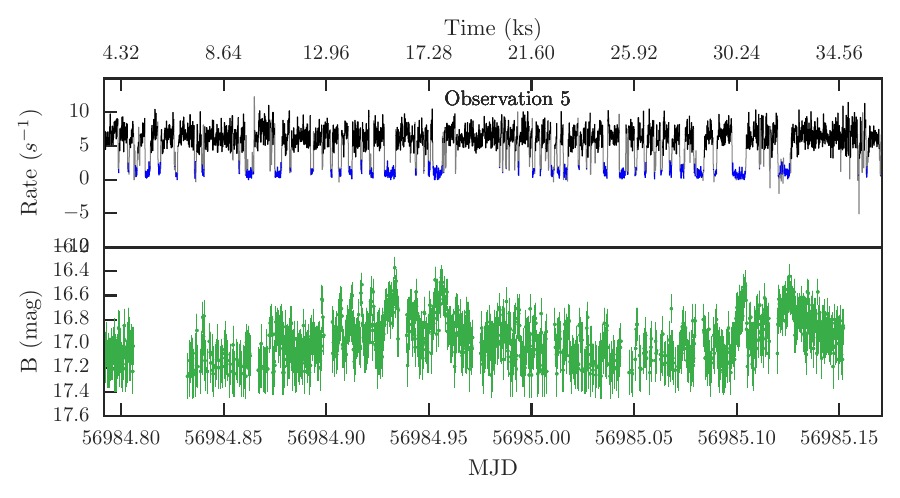}

\vfill
\end{minipage}

\caption{X-ray and optical lightcurves of PSR J$1023$+$0038$  for Obs. 4 and 5 (see
  Table~\ref{table:obssum}). The top panels of each observation show
  the X-ray light curve obtained by adding pn, MOS1, and MOS2 data
  from {\it XMM-Newton} (each background subtracted and exposure
  corrected). Black, blue and red colours are used to depict high, low and
  flare mode, respectively.  The bottom panel in each observation
  represents the {\it XMM-Newton} \textit{B}-filter optical lightcurve.
  Note that, since the {\it XMM-Newton} OM observations start earlier than
  {\it XMM-Newton} EPIC-pn, the start times in above plots are limited by Epic-pn.}

\label{fig:lc}
\end{figure*}

\renewcommand{\thefigure}{\arabic{figure} (Cont.)}
\addtocounter{figure}{-1}

\begin{figure*}
\centering
\begin{minipage}{\textwidth}
\includegraphics[width=\textwidth]{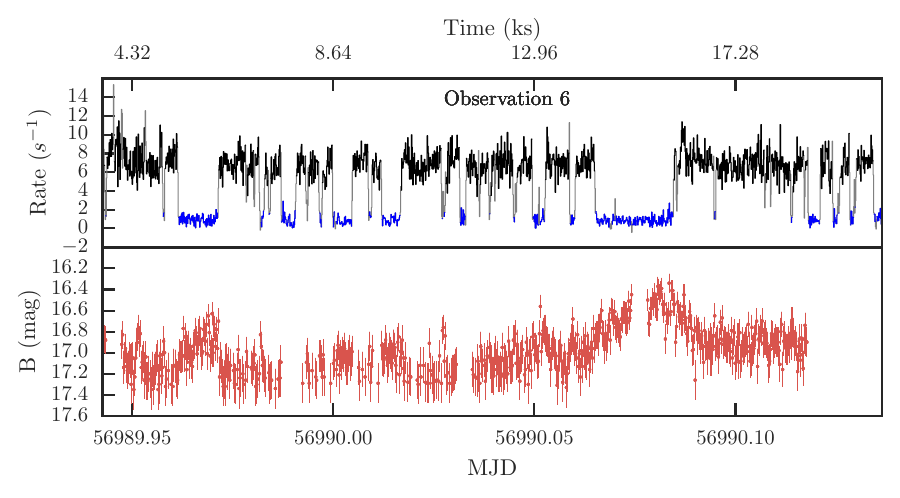}
\includegraphics[width=\textwidth]{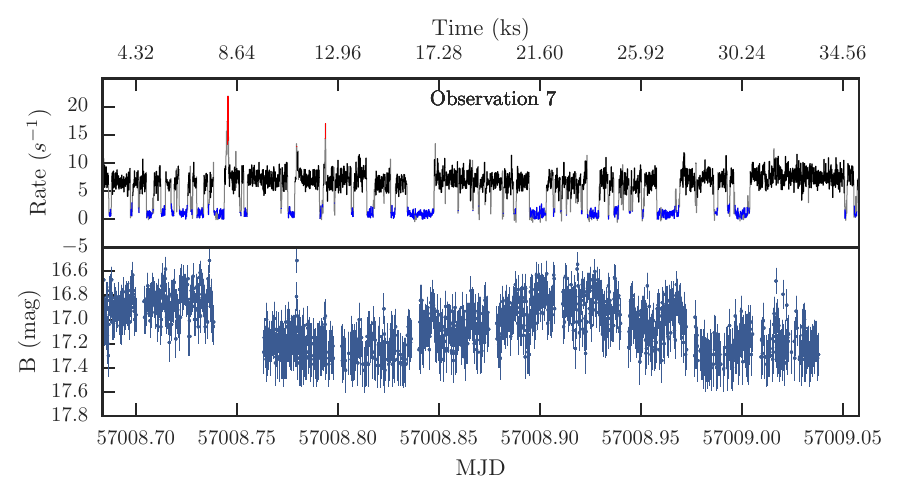}
\vfill
\end{minipage}

\caption{X-ray and optical lightcurves of PSR J$1023$+$0038$  for Obs. 6 and 7 (see
  Table~\ref{table:obssum}).  The top panels of each observation show
  the X-ray light curve obtained by adding pn, MOS1, and MOS2 data
  from {\it XMM-Newton} (each background subtracted and exposure
  corrected). Black, blue and red colours are used to depict high, low and
  flare mode, respectively.  The bottom panel in each observation
  represents the {\it XMM-Newton} \textit{B}-filter optical lightcurve.
  Note that, since the {\it XMM-Newton} OM observations start earlier than
  {\it XMM-Newton} EPIC-pn, the start times in above plots are limited by Epic-pn.}
\end{figure*}

\renewcommand{\thefigure}{\arabic{figure} (Cont.)}
\addtocounter{figure}{-1}

\begin{figure*}
\centering
\begin{minipage}{\textwidth}
\includegraphics[width=\textwidth]{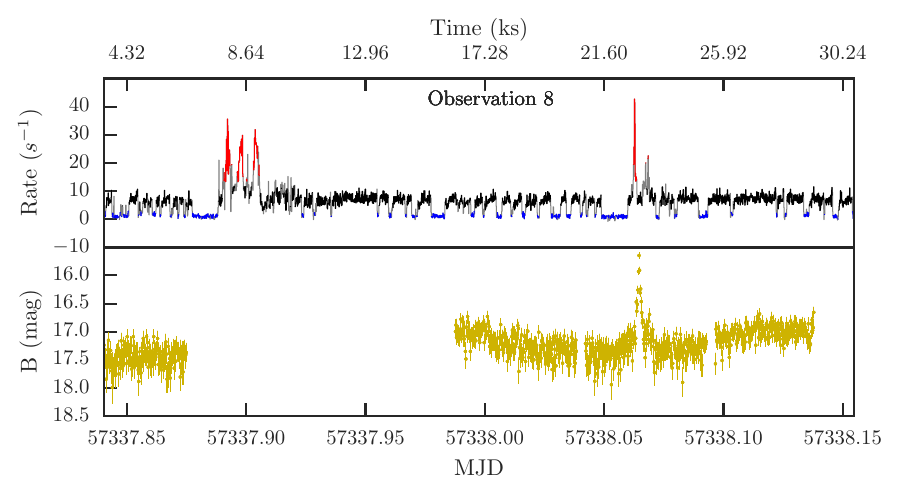}
\includegraphics[width=\textwidth]{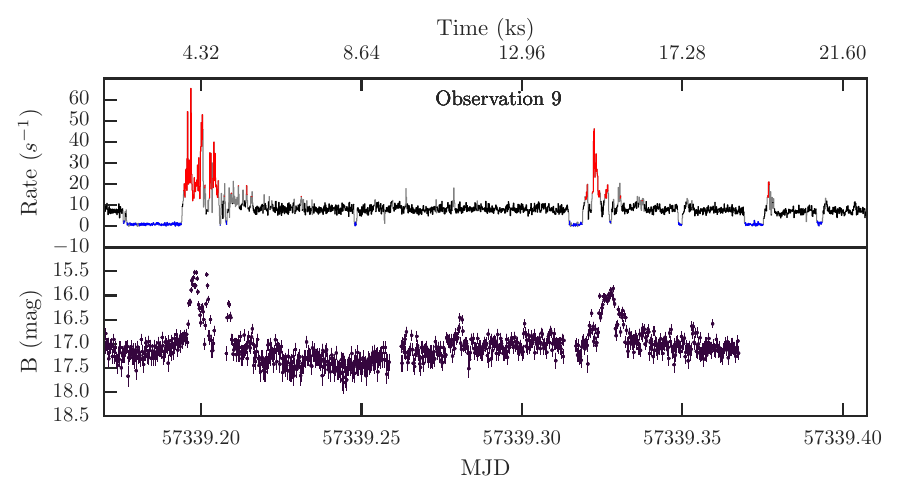}
\vfill
\end{minipage}

\caption{X-ray and optical lightcurves of PSR J$1023$+$0038$ for Obs. 8 and 9 (see
  Table~\ref{table:obssum}).  The top panels of each observation show
  the X-ray light curve obtained by adding pn, MOS1, and MOS2 data
  from {\it XMM-Newton} (each background subtracted and exposure
  corrected). Black, blue and red colours are used to depict high, low and
  flare mode, respectively.  The bottom panel in each observation
  represents the {\it XMM-Newton} \textit{B}-filter optical lightcurve.
  Note that, since the {\it XMM-Newton} OM observations start earlier than
  {\it XMM-Newton} EPIC-pn, the start times in above plots are limited by Epic-pn.}
\end{figure*}

\renewcommand{\thefigure}{\arabic{figure} (Cont.)}
\addtocounter{figure}{-1}

\begin{figure*}
\centering
\begin{minipage}{\textwidth}
\includegraphics[width=\textwidth]{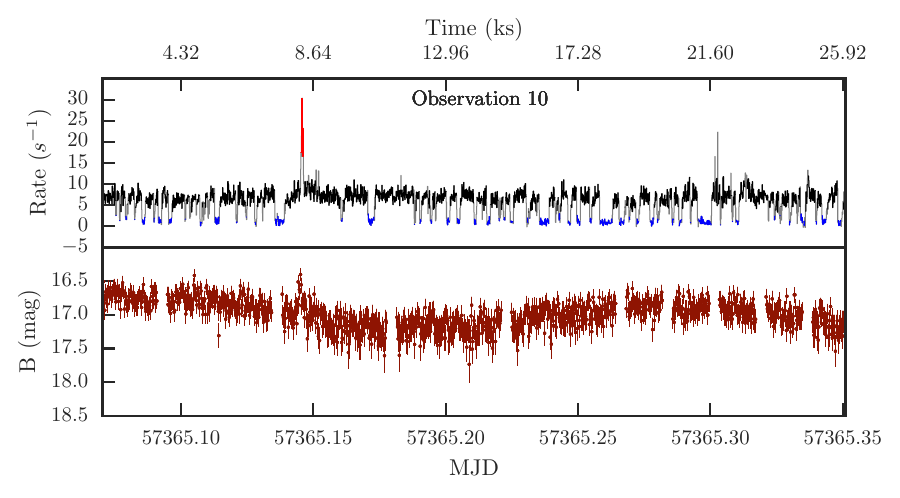}
\vfill
\end{minipage}

\caption{X-ray and optical lightcurves of PSR J$1023$+$0038$  for Obs. 10 (see
  Table~\ref{table:obssum}).  The top panel shows
  the X-ray light curve obtained by adding pn, MOS1, and MOS2 data
  from {\it XMM-Newton} (each background subtracted and exposure
  corrected). Black, blue and red colours are used to depict high, low and
  flare mode, respectively.  The bottom panel in each observation
  represents the {\it XMM-Newton} \textit{B}-filter optical lightcurve.
  Note that, since the {\it XMM-Newton} OM observations start earlier than
  {\it XMM-Newton} EPIC-pn, the start times in above plots are limited by Epic-pn.}
\end{figure*}
\end{document}